\documentclass[final,3p,times,11pt]{elsarticle}
\biboptions{sort&compress}
\usepackage[american]{babel}
\usepackage{tabularx}
\usepackage{multirow}
\usepackage{booktabs}
\usepackage{amsmath}
\usepackage{url}
\usepackage{amssymb}
\usepackage{algorithm}
\usepackage{makecell}
\usepackage{algorithmicx}
\usepackage{algpseudocode}
\usepackage{graphicx}

\usepackage{listings}
\usepackage[section]{placeins}
\usepackage{xspace}
\usepackage{color}
\usepackage{xcolor}
\usepackage{colortbl}
\usepackage{subcaption}
\usepackage[figuresright]{rotating}
\usepackage{tikz}
\usetikzlibrary{positioning,arrows.meta,bending}
\usepackage{cancel}
\usepackage[hidelinks]{hyperref}
\usepackage{xurl}
\usepackage[normalem]{ulem}

\newcommand{\paperpdftitle}{ARGOS: Reinforcement Learning-Driven Multidimensional Elasticity for Service Orchestration in the Computing Continuum}
\newcommand{\papertitle}{ARGOS: Reinforcement Learning-Driven Multidimensional Elasticity for Service Orchestration in the Computing Continuum}
\hypersetup{
  pdftitle={\paperpdftitle},
  pdfauthor={Javier Mateos-Bravo, Sergio Laso, Juan Luis Herrera, Ilir Murturi, Pantelis Frangoudis, Schahram Dustdar}
}


\graphicspath{ {figures/} }

\journal{Internet of Things}

\begin{document}

\begin{frontmatter}

\title{\papertitle}

\author[uex]{Javier Mateos-Bravo\corref{cor1}}

\author[uex,gpi]{Sergio Laso}
\author[uex]{Juan Luis Herrera}
\author[unipr]{Ilir Murturi}
\author[tuwien]{\\Pantelis Frangoudis}
\author[tuwien,icrea]{Schahram Dustdar}

\address[uex]{Department of Computer Science and Telematics Engineering,
              University of Extremadura, Spain}
\address[gpi]{Global Process and Product Improvement S.L., Spain}
\address[unipr]{Department of Mechatronics, University of Prishtina, Kosova}
\address[tuwien]{Distributed Systems Group, TU Wien, Austria}
\address[icrea]{ICREA, Barcelona, Spain}

\cortext[cor1]{Corresponding author. Email: javiermb@unex.es}

\begin{abstract}
Data-intensive services in the Computing Continuum must balance analytics quality, resource usage, and cost across heterogeneous nodes with limited and uneven capacity. This balance becomes especially difficult when resource scaling reaches capacity limits, because changes in demand and cluster pressure must then be absorbed without violating client-defined quality ranges. Existing orchestrators mainly adapt resources, placements, or replicas, while analytics requirements such as coverage, sample, and freshness remain fixed. This article presents ARGOS, the Adaptive Reinforcement Learning-Driven Governance for Orchestrated Services, an end-to-end controller that formulates multidimensional elasticity as a per-request Markov decision process over analytics quality and cluster pressure, supported by capacity-aware admission. ARGOS is evaluated under controlled workloads and time-varying multi-tenant arrivals on a heterogeneous cluster. Across the controlled scenarios, DQN and PPO improve on the static midpoint in every held-out seed, outperform the reactive threshold in the sustained-pressure aggressive-incident profile, and recover 67--72\,\% of the reward of the independently tuned best-fixed reference. A separate live evaluation reports improvements over the static midpoint under realistic and saturated arrivals, with no recorded coverage, sample, CPU, or memory violations and only rare freshness violations. These results support deep reinforcement learning as an adaptive mechanism for multidimensional elasticity when resource scaling alone is insufficient.
\end{abstract}


\end{frontmatter}

\section{Introduction}
\label{sec:introduction}

% Paragraph 1: Continuum constrains replica elasticity
Data-intensive services increasingly run across the Computing Continuum, a layered fabric in which cloud, fog, edge, and mist resources cooperate to move processing closer to where data is produced while preserving access to larger compute pools~\cite{Bonomi2014FogPlatform,Satyanarayanan2017EdgeEmergence,Casamayor2023EdgeIntelligence}. In this article, the mist tier denotes the far-edge Internet of Things (IoT) devices, smartphones, and embedded sensors that generate or first process the data. This fabric is heterogeneous by design because its nodes differ widely in compute capacity, network conditions, energy profile, and availability. Consequently, elasticity cannot be reduced to adding or removing replicas as in a homogeneous cloud. The original notion of elastic processes already framed adaptation along three axes of resources, cost, and quality~\cite{Dustdar2011ElasticProcesses}. Recent work on adaptive data analytics in the Computing Continuum extends that view to three application-level requirements~\cite{Murturi2025ElasticityIoT,Laso2022ElasticAnalytics}. These requirements are coverage, the fraction of spatial or user partitions included in an analysis, sample, the fraction of available observations processed within those partitions, and freshness, the interval between successive result updates. For instance, in an urban-mobility heatmap service, coverage controls geographic breadth, sample controls observation density within those partitions, and freshness controls result recency. Similar quality-resource trade-offs may arise in other configurable analytics services, although the relevant dimensions and their interpretation are application-dependent. Their accepted ranges therefore form part of the request's Service Level Objective (SLO) and expose explicit trade-offs between analytical quality, CPU load, memory pressure, network traffic, and operational cost.

% Paragraph 2: Multidimensional elasticity becomes operational challenge
The practical problem starts when infrastructure elasticity reaches its limits. On the constrained tiers of the Continuum, physical capacity is finite and cannot be expanded on demand. Horizontal and vertical scaling therefore cease to provide additional capacity once the available nodes reach their limits~\cite{laso2026evaluating,LoridoBotran2014AutoscalingReview, laso2025energy}. An unanticipated demand burst on a capacity-limited edge node can then leave the orchestrator with no additional resources to allocate, while requests already accepted by the system must still satisfy their client-defined SLOs. Multidimensional elasticity provides an additional adaptation space by allowing application-level requirements to vary within explicit client ranges. Existing frameworks for multidimensional elasticity have studied this principle for adaptive analytics~\cite{Murturi2025ElasticityIoT,Laso2022ElasticAnalytics}. For configurable analytics, jointly controlling coverage, sample, and freshness on heterogeneous nodes while accounting for observed resource pressure and per-request ranges remains an operational challenge.

% Paragraph 3: Three gaps define controller requirements
Three gaps prevent existing orchestration work from closing this challenge. The first gap (G1) concerns the control space: much reinforcement learning (RL) work for systems optimizes placement, scheduling, replica count, or batching rather than the analytics requirements embedded in per-request contracts~\cite{Mao2016DeepRM,Mao2019Decima,GariRLAutoscalingSurvey2021}. The second gap (G2) concerns evaluation. Online reward traces may still include exploration and parameter updates, so they do not necessarily represent the frozen policy that would be deployed~\cite{Henderson2018DeepRLMatters,Rossi2023DynamicThresholds}. Comparisons with non-learning alternatives must also use the same evaluation conditions. The third gap (G3) concerns reproducibility: evaluations frequently rely on single seeds, uncontrolled hyperparameter changes, or incomplete reporting of experimental settings and artifacts despite repeated calls for sufficient seed counts, explicit variability reporting, and auditable evidence~\cite{Henderson2018DeepRLMatters,Colas2018HowManySeeds,Agarwal2021Precipice,Pineau2021Reproducibility}. Accordingly, G1 requires a controller that adjusts coverage, sample, and freshness while observing resource pressure in the same decision step. G2 requires a protocol that trains each learned policy before evaluating the frozen policy artifact without exploration or parameter updates, alongside non-learning policies under the same conditions. G3 requires the preservation of policy fingerprints, decision logs, telemetry, and provenance for each comparison.

% Paragraph 4: ARGOS architecture, MDP, agents
This article presents ARGOS, the Adaptive Reinforcement Learning-Driven Governance for Orchestrated Services, an orchestration framework for per-request multidimensional elasticity in the Computing Continuum. Structurally, ARGOS combines capacity-aware admission of analytics requests, per-request action selection, configuration distribution to heterogeneous worker nodes, runtime monitoring, and append-only audit records. Algorithmically, ARGOS represents each admitted request as a Markov decision process (MDP) whose state combines cluster pressure and the current analytics configuration, while its actions adjust coverage, sample, and freshness within the accepted request ranges. This common formulation supports both learning and non-learning policies with a shared state representation, action space, safety constraints, and configuration path, enabling systematic comparison under common orchestration conditions.

% Paragraph 5: Three contributions and claims
The contributions of this article are threefold. (i) An end-to-end orchestration design for multidimensional elasticity on heterogeneous clusters, coupling capacity-aware request admission with per-request control of coverage, sample, and freshness under observed resource pressure. The design constrains target values to client-defined ranges, applies resource-safety guards, and records target and realized configurations separately, and its implementation is released as open-source software. (ii) A reproducible methodology for comparing decision policies under a shared orchestration model. Learned policies are trained before frozen evaluation, learning and non-learning approaches share the same state representation, action space, and operational constraints, and each run retains policy fingerprints, decision logs, reward components, node telemetry, and provenance. (iii) A two-stage empirical evaluation on an urban-mobility heatmap service over Seville using simulated per-user movement trajectories~\cite{LopezPerez2023SmartSeville}. The controlled stage compares tabular Q-learning~\cite{Watkins1992QLearning,SuttonBarto2018RL}, Deep Q-Network (DQN)~\cite{Mnih2015DQN}, and Proximal Policy Optimization (PPO)~\cite{Schulman2017PPO} against the static midpoint and reactive threshold baselines and an independently tuned best-fixed reference across five workload profiles and two runtime modes. The live stage then compares the frozen deep policies with the static midpoint under realistic and saturated request arrivals, separating admission behavior, contract compliance, and deployability from controlled model ranking. The evaluation scenario, scripts, result tables, and figure-generation code are also released with the public artefact.

% Paragraph 6: Article structure and scope
The remainder of this article is organized as follows. Section~\ref{sec:motivation} grounds the operational problem in a city-scale urban-mobility service. Section~\ref{sec:related} reviews multidimensional elasticity in the Computing Continuum, RL controllers for orchestration, and reproducible evaluation practices for learned-versus-non-learning comparisons. Section~\ref{sec:argos} presents the ARGOS architecture, its components, its operational lifecycle, and its decision model. Section~\ref{sec:validation} reports the controlled benchmark, frozen-policy comparison, runtime and resource-pressure analyses, and live-trial validation. Section~\ref{sec:discussion} interprets the evidence and states the limits of the claims. Section~\ref{sec:conclusions} summarizes the findings and outlines future extensions.
\section{Motivational use case}
\label{sec:motivation}

% Paragraph 1: Shared service and concurrent municipal use cases
As a concrete motivating scenario, consider a city that operates a shared analytics service for urban mobility. The service ingests movement trajectories gathered from sources such as public-transport fleets and connected user devices and computes heatmaps of where movement concentrates within a requested geographic scope~\cite{LopezPerez2023SmartSeville}. A single service can support several municipal use cases concurrently, submitted as independent requests by public bodies and private operators. The following four examples illustrate how their requirements differ:
\begin{itemize}
    \item \textbf{Infrastructure planning.} The city planner analyzes long-term mobility patterns to plan new transport lines, bus stops, or pedestrian areas. This use case needs a broad city-wide view (high coverage), yet it can tolerate a heatmap refreshed only every few minutes (a long freshness interval).
    \item \textbf{Event management.} City authorities monitor large public events such as concerts or football matches, where crowds concentrate around one district. A high sample fraction of the local trajectories is needed to reveal the build-up and support the redirection of pedestrian flows.
    \item \textbf{Traffic and emergency management.} The traffic department reacts to forming jams and clears routes for emergency vehicles such as ambulances. The heatmap must be refreshed within seconds (a short freshness interval), because stale information can hide a rapidly changing incident.
    \item \textbf{Passenger-flow optimization.} Public-transport operators increase bus frequency where passengers currently concentrate and redesign routes from historical patterns. Current adjustments require a short freshness interval, whereas longer-term planning benefits from a broad city-wide view (high coverage).
\end{itemize}
These requests run concurrently and impose different coverage, sample, and freshness requirements on the finite resources of the same infrastructure.

% Paragraph 2: Accepted ranges, resource demand, and finite capacity
Each request therefore arrives with an SLO contract that specifies accepted ranges for coverage, sample, and freshness and may also include a cost budget. Moving toward broader coverage, a larger sample fraction, or a shorter freshness interval increases the amount or frequency of processing and therefore raises compute, memory, network demand, and estimated operating cost. The platform can absorb growing demand for a while by distributing work across available nodes, delaying new analyses, or rejecting requests whose minimum requirements cannot be accommodated. Once admitted analyses compete for the finite capacity of the same infrastructure, however, the accepted quality ranges provide an additional adaptation space.

\begin{figure}[!htpb]
  \centering
  \includegraphics[width=0.62\linewidth]{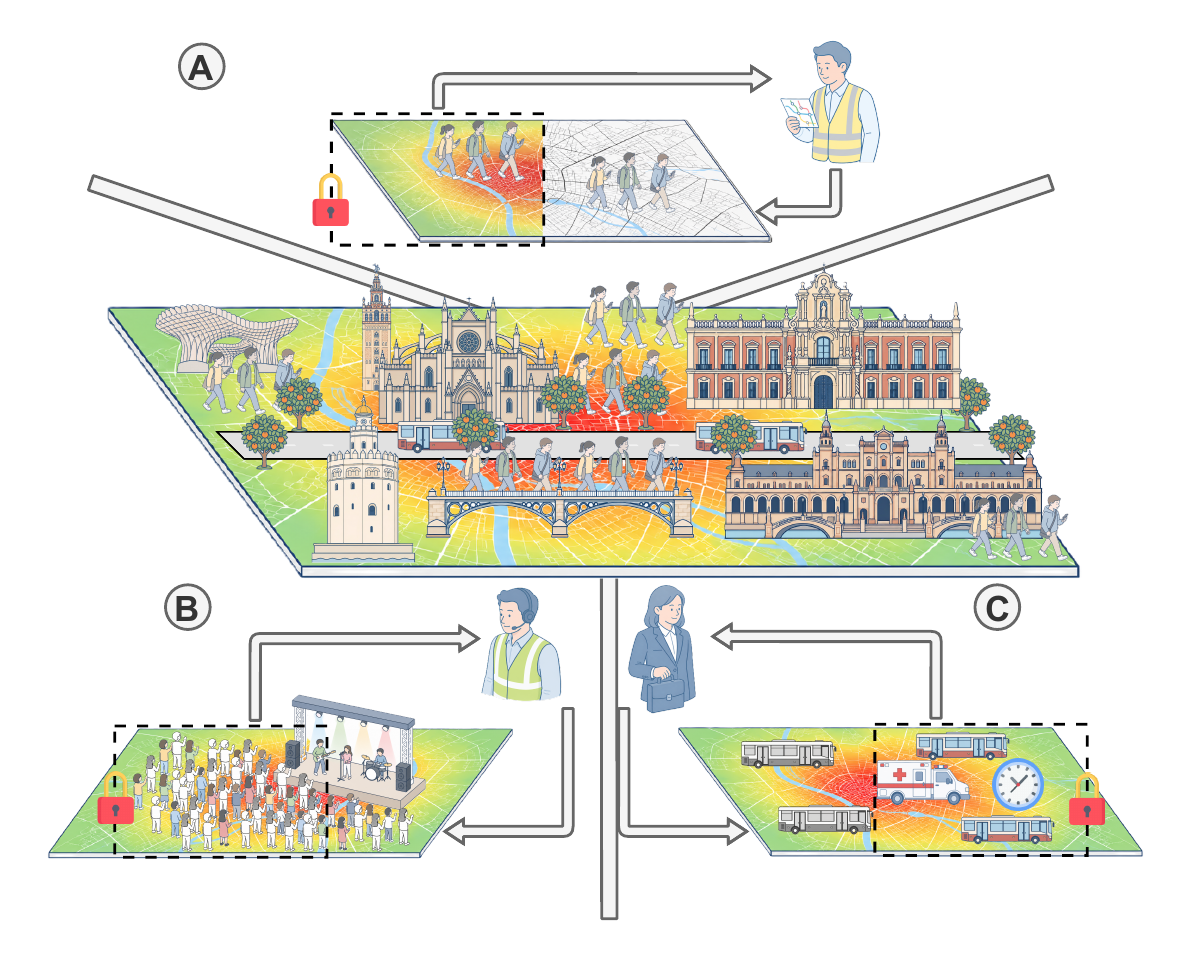}
  \caption{(A) Coverage, (B) sample, and (C) freshness fixed after admission. Padlocks mark what is held fixed in each use case.}
  \label{fig:motivation}
\end{figure}

% Paragraph 3: Fixed operating points prevent adaptation
If the platform selects one operating point at admission and then holds coverage, sample, and freshness fixed for the lifetime of a request, that flexibility cannot be used as shared resource pressure changes. Figure~\ref{fig:motivation} illustrates this problem for three of the use cases above. Panel~(A) represents coverage in infrastructure planning, Panel~(B) represents sample in event management, and Panel~(C) represents freshness in traffic and emergency management. In each case, the corresponding requirement remains fixed after admission even though the conditions under which concurrent analyses share the infrastructure can change. An analysis fixed at a demanding operating point continues consuming scarce capacity when a less demanding configuration would remain within its accepted range. Conversely, an analysis fixed at a conservative operating point cannot use newly available capacity to improve its result. A fixed operating point therefore cannot exploit the flexibility already granted by the client as resource pressure varies.

% Paragraph 4: Motivating requirement and ARGOS response
Within this scenario, the flexibility granted by the accepted ranges motivates an orchestration mechanism that can adjust coverage, sample, and freshness as shared resource pressure changes, while preserving client-defined target bounds and resource-safety constraints. ARGOS addresses this need by treating the three requirements as per-request control dimensions rather than fixed values. This allows each admitted request to move among feasible operating points according to the measured state of the infrastructure. Section~\ref{sec:argos} describes the admission, control, and monitoring mechanisms that implement this behavior.

\section{Related work}
\label{sec:related}

% Paragraph 1: Related-work structure
The literature relevant to ARGOS spans three connected areas. Section~\ref{subsec:related-elasticity} reviews the evolution from infrastructure autoscaling toward application-level and multidimensional elasticity in the Computing Continuum. Section~\ref{subsec:related-rl} examines reinforcement-learning controllers for resource management and orchestration, including scaling, placement, and scheduling. Section~\ref{subsec:related-evaluation} considers the baselines and reproducibility practices used to evaluate learned controllers. The analysis of each area then identifies the limitation that motivates the corresponding design choices in ARGOS.

\subsection{Multidimensional elasticity in the Computing Continuum}
\label{subsec:related-elasticity}

% Paragraph 2: From infrastructure scaling to application-level elasticity
Cloud and edge elasticity initially focused on adapting infrastructure resources to changing demand. Lorido-Botran et al.~\cite{LoridoBotran2014AutoscalingReview} review autoscaling techniques that adjust provisioned resources while balancing service objectives and cost. Casalicchio~\cite{Casalicchio2019Autoscaling} examines the metrics used to trigger container autoscaling and evaluates CPU-based replica adaptation in Kubernetes. Rossi et al.~\cite{Rossi2019HorizVertScaling} extend this resource-oriented view by coordinating horizontal and vertical scaling through reinforcement learning. Other approaches move adaptation closer to application requirements. Trihinas et al.~\cite{Trihinas2015AdaM} adapt monitoring through dynamic sampling and filtering on constrained IoT devices, Khaleq and Ra~\cite{Khaleq2021QoSAwareAutoscaling} couple reinforcement-learning autoscaling with response-time constraints, and Murturi and Dustdar~\cite{Murturi2022DECENT} capture application elasticity requirements for decentralized runtime enforcement at the edge. These works broaden elasticity beyond fixed resource thresholds, but they do not jointly control coverage, sample, and freshness within the accepted ranges of each analytics request.

% Paragraph 3: Multidimensional analytics elasticity
A complementary line treats analytical quality itself as elastic. Laso et al.~\cite{Laso2022ElasticAnalytics} introduce elastic data analytics as a way to adapt analytics behavior to shared and constrained edge infrastructures. Their framework formalizes multidimensional elasticity through coverage, sample, and freshness for analytics services in the Computing Continuum~\cite{Murturi2025ElasticityIoT}. This framework provides the conceptual base for ARGOS. Its testbed characterizes the impact of all three analytics requirements, while its Q-learning study adapts resource capacity, sample, and freshness over homogeneous simulated nodes and leaves coverage outside the learned state and action spaces. ARGOS builds on that foundation by controlling coverage, sample, and freshness with observed resource pressure in one per-request MDP deployed over heterogeneous nodes.

\subsection{Reinforcement-learning controllers for orchestration}
\label{subsec:related-rl}

% Paragraph 4: Learned resource allocation and scheduling
Reinforcement learning has been applied to several resource-management problems in cloud and cluster systems. Tesauro et al.~\cite{Tesauro2006Hybrid} combine reinforcement learning with queueing models for autonomic server allocation, using offline learning before the learned policy controls the system. DeepRM~\cite{Mao2016DeepRM} formulates cluster resource management as a deep reinforcement-learning problem in which jobs with multiple resource demands are scheduled from experience. Decima~\cite{Mao2019Decima} extends learned scheduling to data-processing clusters by optimizing the execution of dependency-aware workloads. These systems demonstrate that learned controllers can support resource allocation and scheduling, but their actions determine resource assignment or execution order rather than the analytics requirements of individual requests.

% Paragraph 5: Learned orchestration on heterogeneous infrastructures
Later work applies reinforcement learning to increasingly heterogeneous orchestration settings. Garí et al.~\cite{Gari2021QLearningAutoscaling} formulate scientific-workflow autoscaling as an MDP and use Q-learning to trade execution time against cloud cost. Goudarzi et al.~\cite{Goudarzi2024DistributedDRL} use distributed deep reinforcement learning for application placement across edge and fog resources. Wang et al.~\cite{WangGoudarzi2024EdgeFogScheduling} implement a learned scheduler in FogBus2 that selects execution nodes while considering response time and load balance. Mampage et al.~\cite{Mampage2023ServerlessRL} apply deep reinforcement learning to function scheduling in a multi-tenant serverless environment and evaluate it on a deployed Kubernetes cluster. These studies extend learned orchestration to heterogeneous and deployed infrastructures, but their control targets remain scaling, placement, or scheduling. ARGOS instead controls coverage, sample, and freshness within client-defined ranges while observing shared resource pressure.

\subsection{Evaluation rigor: baselines and reproducibility}
\label{subsec:related-evaluation}

% Paragraph 6: Comparison policies
The conclusions drawn from a learned controller depend in part on the policies used for comparison. Rossi et al.~\cite{Rossi2023DynamicThresholds} use reinforcement learning to adapt scaling thresholds rather than relying on fixed single-metric thresholds, showing how the reference policy can affect the resulting comparison. Mampage et al.~\cite{Mampage2023ServerlessRL} evaluate their learned serverless scheduler against several scheduling alternatives on a deployed cluster. Garí et al.~\cite{GariRLAutoscalingSurvey2021} survey RL-based autoscaling and report substantial variation in state representations, actions, rewards, and evaluation designs. These works motivate comparisons against several meaningful alternatives rather than against a single simple baseline.

% Paragraph 7: Reproducibility of learned-controller evaluations
Reproducibility introduces a concern. Henderson et al.~\cite{Henderson2018DeepRLMatters} show that deep reinforcement-learning results can vary substantially with random seeds, network architectures, hyperparameters, and implementation choices. Colas et al.~\cite{Colas2018HowManySeeds} study how the number of random seeds affects the reliability of comparisons between stochastic learning algorithms. Agarwal et al.~\cite{Agarwal2021Precipice} show that aggregate conclusions can be unstable with limited runs and propose more robust statistical summaries, while Pineau et al.~\cite{Pineau2021Reproducibility} consolidate reproducibility practices for machine-learning research. For orchestration experiments on heterogeneous hardware, these results motivate separating training from evaluation, retaining seed-level outcomes, and preserving the configuration, decision, and telemetry evidence required to trace each policy.

% Paragraph 8: Closing synthesis and comparative summary
Across the three areas reviewed above, the remaining gap lies in their combination. Multidimensional elasticity establishes coverage, sample, and freshness as adaptable analytics requirements, while RL-based orchestration provides mechanisms for sequential control over shared infrastructure but primarily targets scaling, placement, or scheduling. Evaluation research further shows the importance of meaningful comparison policies, repeated runs, and reproducible experimental evidence. ARGOS combines these elements by controlling all three analytics requirements under observed resource pressure, comparing learned and non-learning policies under the same orchestration model, and separating policy training from frozen evaluation while retaining traceable experimental evidence. Table~\ref{tab:related-comparison} summarizes this positioning through the control target, evaluation setting, and reported comparison design of representative orchestration approaches.

\begin{table*}[!htpb]
\centering
\begingroup
\footnotesize
\setlength{\tabcolsep}{5pt}
\renewcommand{\arraystretch}{1.15}
\begin{tabular}{llll}
\toprule
Study &
\makecell[l]{Approach and control target} &
\makecell[l]{Evaluation setting} &
\makecell[l]{Reported comparison} \\
\midrule
Autonomic allocation~\cite{Tesauro2006Hybrid}
  & \makecell[l]{Hybrid RL, server allocation}
  & \makecell[l]{Web-server resource\\management}
  & \makecell[l]{Model-based allocation} \\

DeepRM~\cite{Mao2016DeepRM}
  & \makecell[l]{Deep RL, cluster resources}
  & Simulation
  & \makecell[l]{Resource-management\\heuristics} \\

Decima~\cite{Mao2019Decima}
  & \makecell[l]{RL, dataflow scheduling}
  & \makecell[l]{Data-processing cluster}
  & \makecell[l]{Scheduling alternatives} \\

Workflow autoscaling~\cite{Gari2021QLearningAutoscaling}
  & \makecell[l]{Q-learning, workflow\\autoscaling}
  & \makecell[l]{Cloud workflow evaluation}
  & \makecell[l]{Autoscaling alternatives} \\

Application placement~\cite{Goudarzi2024DistributedDRL}
  & \makecell[l]{Distributed deep RL, IoT\\placement}
  & \makecell[l]{Simulation and testbed}
  & \makecell[l]{Placement alternatives} \\

Dynamic thresholds~\cite{Rossi2023DynamicThresholds}
  & \makecell[l]{RL-tuned thresholds,\\application scaling}
  & \makecell[l]{Simulation and prototype}
  & \makecell[l]{Scaling policies} \\

Continuum scheduling~\cite{WangGoudarzi2024EdgeFogScheduling}
  & \makecell[l]{Deep RL, edge/fog scheduling}
  & \makecell[l]{FogBus2 testbed}
  & \makecell[l]{Scheduling alternatives} \\

Serverless RL~\cite{Mampage2023ServerlessRL}
  & \makecell[l]{Deep RL, serverless scheduling}
  & \makecell[l]{Kubernetes cluster}
  & \makecell[l]{Baseline schedulers} \\

\midrule
ARGOS, this work
  & \makecell[l]{Per-request MDP, analytics\\quality}
  & \makecell[l]{Heterogeneous cluster}
  & \makecell[l]{Midpoint, threshold, and\\best-fixed} \\
\bottomrule
\end{tabular}
\endgroup
\caption{Representative orchestration approaches and evaluation designs.}
\label{tab:related-comparison}
\end{table*}
\section{ARGOS}
\label{sec:argos}

\begin{figure*}[!htpb]
  \centering
  \includegraphics[width=\linewidth]{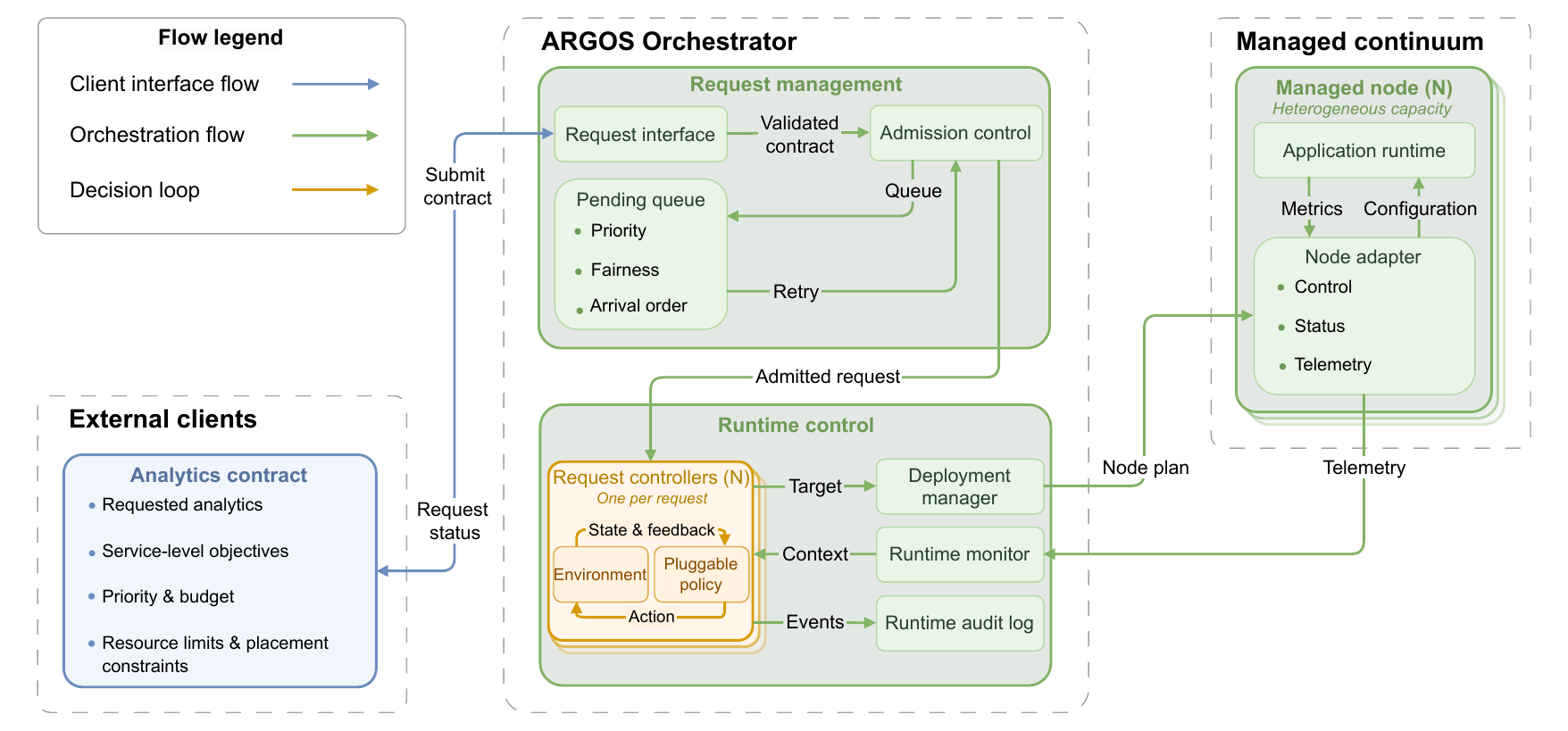}
  \caption{ARGOS deployment architecture and runtime interfaces.}
  \label{fig:architecture}
\end{figure*}

% Paragraph 1: Deployment architecture and boundaries
ARGOS is a deployment-level orchestration framework that mediates between tenant applications and analytics runtimes distributed across the Computing Continuum. Figure~\ref{fig:architecture} separates this deployment into three architectural domains: external clients that express requests as analytics contracts, the ARGOS orchestrator that admits and controls those requests, and the managed Continuum whose nodes execute the applications. Using the current runtime context, the orchestrator selects values for coverage, sample, and freshness within the ranges accepted by each request and translates them into placement and configuration instructions, while every managed node exposes a uniform control and telemetry interface. This separation keeps the orchestration logic independent of a particular analytics implementation and confines service-specific behavior to the node-side runtime adapter.

% Paragraph 2: General architecture and case-study mapping
The architecture supports configurable analytics services through a common service interface. The urban-mobility heatmap described in Section~\ref{sec:motivation} is the current prototype instantiation of this interface. A service integrated with ARGOS must accept target quality values, execute through the placement selected by the orchestrator, and report realized quality and resource telemetry during execution. In the current prototype, each assigned node processes one trajectory partition and produces one heatmap shard. Target coverage determines the desired number of partitions, sample controls the fraction of trajectory observations processed within each partition, and freshness controls the update interval. Section~\ref{subsec:components} details the three domains in Figure~\ref{fig:architecture}, Section~\ref{subsec:flows} describes the operational lifecycle, and Section~\ref{subsec:mdp} formalizes the decision model. The ARGOS source code, configuration files, evaluation scripts, and reproducibility artefacts are publicly available on Zenodo~\cite{MateosBravo2026ARGOS}.

\subsection{Components}
\label{subsec:components}

% Paragraph 3: External clients and request interface
\textbf{External clients} comprise tenant applications, operators, and services that request analytics without depending on the internal controller. They submit an \textbf{Analytics contract} through the \textbf{Request interface}, the public application programming interface (API) that also exposes request and cluster status. The contract carries accepted ranges for coverage, sample, and freshness together with the tenant identity, priority, resource caps, and optional cost and placement constraints. This public interface is the architectural entry point and is also used by experimental workload generators acting as ordinary external clients during validation.

% Paragraph 4: ARGOS orchestrator
The \textbf{ARGOS orchestrator} coordinates request admission, per-request control, and deployment across the managed Continuum. It maintains the cluster view, request lifecycle, admission state, per-request controllers, and the node plans sent to managed nodes. Although one active orchestrator is sufficient for the reference deployment, the role is not tied to a specific machine. In distributed deployments, peer heartbeats, capacity-aware leader election, and a message bus allow another eligible node to assume the active orchestrator role.

% Paragraph 5: Admission control and pending queue
The \textbf{Admission control} component validates each contract and determines whether its minimum requirements can be placed under the current capacity and placement limits. A structurally infeasible request is rejected, and the Request interface returns the rejection status and reason to the client. A contract is also structurally infeasible when no number of active nodes yields a realized coverage inside its accepted range. A feasible request with available capacity is admitted, whereas a temporarily blocked request enters the \textbf{Pending queue} and is returned with its current queue position. Queued requests are ordered by priority, tenant fairness debt, and arrival order, then reconsidered when capacity changes. Admission therefore protects the managed infrastructure before any policy begins to adapt an accepted request.

% Paragraph 6: Per-request controller
Each admitted request receives an independent \textbf{Per-request controller}. This design keeps the control state and policy history of each request separate, so the state or learning history of one request is not used directly to select actions for another. The controller combines the \textbf{Environment} and \textbf{Pluggable policy} roles shown together in Figure~\ref{fig:architecture}. The environment converts contract position, cluster telemetry, service measurements, and admission context into the state, masks ineffective or unsafe actions, applies the selected adjustment with a final range clamp, and computes its reward. Its policy then maps that state to an action and may be deterministic or learned, but every alternative uses the same state, actions, masking rules, and configuration path described in Section~\ref{subsec:mdp}.

% Paragraph 7: Monitoring, placement, and runtime audit
The \textbf{Runtime monitor} polls the node adapters and aggregates CPU, memory, bandwidth, latency, realized quality, and request-lifecycle information into the context consumed by each controller. The \textbf{Deployment manager} converts target coverage into a desired node count, selects nodes with available capacity, and distributes the complete target configuration. With $N$ active nodes, realized coverage is the assigned-node count $k$ divided by $N$. The \textbf{Deployment manager} first determines the node counts whose realized coverage $k/N$ lies inside $[c_{\min},c_{\max}]$ and respects the request's placement limits. The desired count is $\lceil \mathrm{coverage}\cdot N \rceil$ restricted to this feasible set. This discrete mapping can differ from the continuous target, especially in a small cluster, so the \textbf{Runtime audit log} records both values together with lifecycle transitions, decisions, configurations, telemetry, violations, and errors. Runtime auditing is part of ARGOS because an operator needs traceability, whereas campaign aggregation and figure generation remain external validation tasks.

% Paragraph 8: Managed continuum and application adapter
The \textbf{Managed Continuum} contains heterogeneous nodes that expose the same \textbf{Node adapter} to the orchestrator. Each adapter applies versioned node plans, supervises one or more \textbf{Application runtimes}, and reports host and service telemetry through its probe. The architectural role of an application runtime is generic, but the current prototype implements a geographic heatmap adapter. That adapter partitions trajectories by assigned node, applies the selected sample and freshness values, and reports processing time, data volume, spatial fidelity, and update age. Thread and process execution are implementation modes of this adapter rather than separate architectural components.

\subsection{Operational lifecycle}
\label{subsec:flows}

% Paragraph 9: Runtime paths and bootstrap
The architecture exposes two runtime paths over the life of a request, drawn in Figure~\ref{fig:architecture} as its flow legend. The request path follows the contract-submission and orchestration flows: it validates, admits, queues, or rejects a submitted contract. The adaptation path corresponds to the decision loop inside each request controller together with the orchestration flows around it: it repeatedly observes an admitted request and updates its operating point. Before either path becomes active, the orchestrator registers the managed nodes, checks their health, and records the initial capacity view. Request processing begins only after one active orchestrator and the required node adapters are ready.

% Paragraph 10: Request contract
The request path starts with tenants such as those behind the use cases of Section~\ref{sec:motivation}. Each contract carries ranges $[c_{\min}, c_{\max}]$, $[\sigma_{\min}, \sigma_{\max}]$, and $[\phi_{\min}, \phi_{\max}]$ for coverage, sample, and freshness. It also supplies the tenant identity, a critical, standard, or best-effort priority, an optional cost budget, per-request CPU and memory caps, and placement limits on node count and concurrent runtimes per node. These fields provide the complete information needed to validate the request, order it fairly, and determine whether its minimum operating point is feasible.

% Paragraph 11: Admission and initial placement
The request path carries the contract from its source to a placement decision. After validation, ARGOS initializes a target configuration within the accepted ranges, using their midpoints by default. The orchestrator then computes the smallest node count whose realized coverage satisfies the contract and applies one of three transitions: immediate assignment when capacity exists, temporary queueing when the request remains feasible, or rejection when no node count keeps realized coverage inside the accepted range under the active nodes and placement limits. Every transition is written to the runtime audit log, which preserves the complete admission history for operational inspection and later validation.

% Paragraph 12: Adaptation path
The adaptation path repeats at the polling cadence and turns the latest runtime context into the next operating point. ARGOS collects node and service telemetry, updates request context, encodes the MDP state, and builds the admissible action set. The mask excludes actions that would have no effect because the corresponding target is already at a contract bound, as well as actions that would increase load when placement capacity is exhausted or the mean observed CPU or memory usage approaches the request's cap. The active controller therefore selects only actions that can be applied, after which the environment updates the target configuration and the placement component distributes a new node plan when required. At a configurable polling cadence, the loop checks elapsed time before decision and distribution against an iteration budget. If the budget is exhausted, the next phase is skipped and the event is recorded. The audit log stores the state, action, target and realized configuration, reward components, algorithm metadata, and operational errors for the path summarized in Algorithm~\ref{alg:control-loop}.

% Paragraph 13: Policy lifecycle outside the deployed architecture
Policy training is a lifecycle activity. ARGOS can update a policy in training mode, serialize the resulting artifact, and later load it in deployment or evaluation mode with exploration and parameter updates disabled. The deployed architecture only requires the policy interface inside the per-request controller. The training budget, artifact freezing procedure, comparator selection, and provenance checks belong to the validation protocol and are specified in Section~\ref{subsec:mdp} and Section~\ref{sec:validation}.

\subsection{Decision model}
\label{subsec:mdp}

% Paragraph 14: MDP state, thirteen dimensions
Each per-request controller represents its request as an MDP over an aggregated operational state. The state at iteration $t$ is encoded as
\begin{equation*}
  s_t =
  \bigl(
  \operatorname{cpu}_t,
  \operatorname{mem}_t,
  \operatorname{bw}_t,
  c_t,
  \sigma_t,
  \phi_t,
  d_t,
  x_t,
  j_t,
  h_t,
  u_t,
  \ell_t,
  v_t
  \bigr),
\end{equation*}
where $\operatorname{cpu}_t$, $\operatorname{mem}_t$, and $\operatorname{bw}_t$ are observed resource-pressure buckets. The terms $c_t$, $\sigma_t$, and $\phi_t$ encode the current position of coverage, sample, and freshness relative to the accepted range of the request. The service-duty bucket $d_t$ relates processing time to the requested refresh interval, while $x_t$ encodes the offered input-load level. The remaining terms encode active jobs $j_t$, free capacity $h_t$, pending demand $u_t$, latency $\ell_t$, and recent SLO violations $v_t$. The latter counts decision epochs with any contract or resource violation over a configurable rolling window of length $W$. The prototype fixes $W=20$ to retain recent violations across several decisions without accumulating the full request history. This value is fixed a priori and is not tuned during evaluation. Resource and runtime dimensions use non-uniform cuts chosen around operational pressure regions, from idle through warning to overload. Contract positions are canonicalized, normalized to $[0,1]$, and divided at $\{0.2,0.4,0.6,0.8\}$, which prevents two arithmetically equivalent operating points from entering different state buckets. The resulting 13-dimensional state is used directly by Q-learning and vectorized for DQN and PPO. The rolling count summarizes the recent history but does not expose the order of the $W$ binary events to the agent. The model is therefore an operational MDP approximation over the telemetry available to the controller rather than a claim of complete observability.

\begin{algorithm}[!h]
\begin{algorithmic}[1]
\Require Polling interval $\Delta$, iteration budget $B$, active requests $\mathcal{R}_t$ each with policy $\pi_r$, mode $m_r$, environment $E_r$, and target configuration $\boldsymbol{\theta}_t^r$, and node set $\mathcal{N}$.
\Loop
  \State $t_0 \gets \Call{now}{}$
  \State $\mathbf{m} \gets \Call{PollMetrics}{\mathcal{N}}$
  \If{$\Call{now}{} - t_0 < B$}
    \State $\mathcal{P} \gets \emptyset$ \Comment{node plans}
    \ForAll{$r \in \mathcal{R}_t$}
      \State $s_t^r \gets E_r.\Call{EncodeState}{\mathbf{m},\, \text{ctx}_r}$
      \State $\mathcal{A}_t^r \gets E_r.\Call{AvailableActions}{}$
      \State $a_t^r \gets \pi_r.\Call{SelectAction}{s_t^r,\, \mathcal{A}_t^r}$
      \State $(s_{t+1}^r,\, \boldsymbol{\theta}_{t+1}^r,\, \rho_t^r,\, \text{comp}_t^r)
      \gets E_r.\Call{Apply}{a_t^r}$
      \If{$m_r = \textsc{train}$}
        \State $\pi_r.\Call{Update}{s_t^r,\, a_t^r,\, \rho_t^r,\, s_{t+1}^r}$
      \EndIf
      \State $\Call{Persist}{r,\, s_t^r,\, a_t^r,\, \boldsymbol{\theta}_{t+1}^r,\, \rho_t^r,\, \text{comp}_t^r}$
      \If{$\boldsymbol{\theta}_{t+1}^r \neq \boldsymbol{\theta}_t^r$}
        \State $\mathcal{P} \gets \Call{UpdatePlans}{\mathcal{P},\, r,\, \boldsymbol{\theta}_{t+1}^r}$
      \EndIf
    \EndFor
    \If{$\mathcal{P} \neq \emptyset$}
      \If{$\Call{now}{} - t_0 < B$}
        \State $\Call{Distribute}{\mathcal{P},\, \mathcal{N}}$
      \Else
        \State $\Call{RecordBudgetEvent}{}$
      \EndIf
    \EndIf
  \Else
    \State $\Call{RecordBudgetEvent}{}$
  \EndIf
  \State $t_1 \gets \Call{now}{}$
  \State $\Call{Sleep}{\max(0,\, \Delta - (t_1 - t_0))}$
\EndLoop
\end{algorithmic}
\caption{ARGOS runtime orchestration loop.}
\label{alg:control-loop}
\end{algorithm}

% Paragraph 15: Seven masked actions
The action space contains seven actions:
\begin{equation*}
  \mathcal{A} =
  \{\operatorname{hold},
  \operatorname{inc}_c,
  \operatorname{dec}_c,
  \operatorname{inc}_{\sigma},
  \operatorname{dec}_{\sigma},
  \operatorname{inc}_{\phi},
  \operatorname{dec}_{\phi}\}.
\end{equation*}
Coverage and sample actions move the corresponding target value by $0.05$. Freshness actions change the target interval by 10 seconds, with $\operatorname{dec}_{\phi}$ producing fresher updates and $\operatorname{inc}_{\phi}$ relaxing the requirement. Before selecting an action, ARGOS masks actions that would have no effect because the corresponding target is already at a contract bound, together with actions that would increase load when placement capacity is exhausted or the mean observed CPU or memory usage approaches the request's cap. A final clamp remains as a defensive check, so no target selected by an applied action can leave the accepted range.

% Paragraph 16: Reward, quality minus penalties
Each action is scored by a reward that is a clipped linear score over quality and penalties:
\begin{equation*}
  r_t =
  \operatorname{clip}_{[-1,1]}
  \left(
  q_t
  - p^{\operatorname{resource}}_t
  - p^{\operatorname{cost}}_t
  - p^{\operatorname{range}}_t
  - p^{\operatorname{overload}}_t
  - p^{\operatorname{fairness}}_t
  - p^{\operatorname{latency}}_t
  - p^{\operatorname{capacity}}_t
  \right).
\end{equation*}
The quality term $q_t$ is the mean of three normalized scores: coverage and sample are better near the upper end of the accepted range, while freshness is better near the lower end because lower intervals produce more recent analytics. Seven penalties price the operational state of the request. The resource term combines observed CPU, memory, and bandwidth pressure with normalized service duty, using the larger available signal and falling back to requirement-induced pressure when neither is available. Cost follows the estimate induced by coverage, sample, and freshness, while the range penalty independently checks coverage, sample, and freshness against their accepted ranges and penalizes any out-of-range target value. Overload activates when observed CPU or memory exceeds its hard cap, fairness reflects tenant debt, latency captures response-time excess, and capacity captures waiting demand when no placement remains. Quality and cost are evaluated from the updated target configuration, whereas resource penalties use the latest telemetry collected before action selection.

% Paragraph 17: Fixed reward weights
Three fixed weights anchor the aggregator: quality carries $1.0$, resource pressure carries $1.2$, and cost carries $0.5$. Resource and cost are normalized into a shared unit penalty budget whose maximum combined contribution is $1.0$, with the larger resource coefficient giving resource pressure more influence than cost within that blend. Overload is computed separately for CPU and memory, with each resource contributing at most $0.5$ and their joint maximum therefore equal to $1.0$. The remaining maximum contributions are $0.5$ for the three range checks together, $0.35$ for latency, $0.30$ for capacity, and $0.20$ for fairness. These limits place the unclipped score between $-3.35$ and $1.0$ before clipping it to $[-1,1]$, while persistence of every component keeps the original composition auditable. In the evidence retained for this study, the clipping bound is never active: none of the 389{,}376 training, controlled-evaluation, and live decisions is clipped, and the observed unclipped score remains within $[-0.846,0.689]$.

% Paragraph 18: Three learned controllers
Three learned controllers are instantiated against the same MDP. The \textbf{Q-learning} controller stores values by encoded state and action index, initializes every unvisited action optimistically at $3.0$, and resolves ties through a private seeded generator rather than favoring the first action. Its learning rate is $0.1$, its discount factor is $0.95$, and epsilon decreases from $0.15$ to $0.01$ over the configured training horizon. The \textbf{DQN} controller replaces the table with a neural value approximator of hidden size $64$ and combines a replay buffer of $10{,}000$ transitions, a 32-transition warm-up, the Adam optimizer, a smooth L1 Bellman loss, and a target network synchronized every 25 steps. The \textbf{PPO} controller uses an actor-critic of the same hidden size, collects 32-step rollouts, estimates generalized advantages with $\lambda=0.95$, and performs four clipped-update epochs with minibatches of 16. Its clipping ratio is $0.2$, value weight is $0.5$, and entropy weight is $0.01$~\cite{Schulman2017PPO}. All controllers use private seeded generators, while PPO samples from its masked categorical policy during training and uses masked argmax during frozen evaluation.

% Paragraph 19: Comparators, training, frozen evaluation
The comparison uses two untuned deployable baselines and one tuned static reference. \textbf{Static midpoint} holds the midpoint configuration through the same per-request selector and environment as the learned agents, whereas \textbf{Threshold} reacts whenever CPU, memory, or service duty reaches its warning region, demand queues, or the recent SLO window contains a violation. Under pressure, Threshold cycles through sample reduction, coverage reduction, and freshness relaxation, then applies the inverse quality improvements after pressure subsides. The \textbf{Best-fixed} reference holds one of nine predeclared configurations formed by the eight contract vertices and their joint midpoint, selected independently for each workload profile through tuning seeds $\{101,202,303\}$. Training uses disjoint seeds $\{11,22,33,44,55\}$, while frozen evaluation uses held-out seeds $\{111,222,333,444,555\}$. Each learned controller trains for 2048 one-second decisions under the four-phase input-load schedule $\{2,16,64,4\}$, with 16 decisions per phase and 32 complete cycles, before its serialized policy is loaded for 64 decisions under the distinct input-load schedule $\{3,12,48,6\}$ with exploration and updates disabled. This separation changes the random stream and offered-load trace between training and evaluation. The campaign records the SHA-256 hash and parameter fingerprint of every policy artifact, rejecting any run whose loaded fingerprint, iteration count, runtime, source revision, or provenance manifest does not match the declared protocol.

\section{Experimental validation}
\label{sec:validation}

% Paragraph 1: Two validation stages
This section validates ARGOS on the analytics service and decision model of Section~\ref{sec:argos} through a two-stage protocol. The controlled stage compares every frozen learned policy with the three non-learning comparators, the static midpoint, the reactive threshold, and the tuned \textit{best-fixed} reference, under identical load schedules, and answers the model-ranking question. The live stage introduces realistic and saturated request arrivals, and answers the deployability and contract-safety question without conflating it with model ranking. Section~\ref{subsec:setup} details the testbed, the workload profiles, and the campaign protocol that both stages share, Section~\ref{subsec:bench-results} reports the controlled comparison, Section~\ref{subsec:resources} analyzes the quality-against-pressure trade that the comparison exposes, and Section~\ref{subsec:live} reports the live trials.

\subsection{Experimental setup}
\label{subsec:setup}

% Paragraph 2: Testbed and dataset
The reference deployment is a cluster of three heterogeneous nodes, summarized in Table~\ref{tab:testbed-nodes}. The Large node coordinates the cluster and also executes analytics, while the Medium and Small nodes provide worker capacity under progressively tighter resource limits. This asymmetry is intentional and represents a common Continuum deployment in which a more capable node coordinates smaller worker nodes~\cite{Bonomi2014FogPlatform,Satyanarayanan2017EdgeEmergence}. The validation pins the active orchestrator to the Large node, a role that Section~\ref{subsec:components} shows is otherwise mobile. The cluster runs on virtualized cloud resources, which supports reproduction, while the node profiles emulate continuum heterogeneity from 1 to 16 virtual CPUs (vCPUs). The service is instantiated with a Seville urban-mobility dataset of 60 per-user movement trajectories generated with the Opportunistic Network Environment mobility simulator~\cite{Keranen2009ONE}, together containing 128{,}103 timestamped location points, representative of smart-city mobility analytics~\cite{LopezPerez2023SmartSeville}. Each controlled campaign fixes one runtime mode, and the comparison is repeated on the thread and process runtimes, so that each controlled comparison is repeated under both concurrency models.

% Paragraph 3: Workload profiles
The controlled stage exercises every controller on five recurrent workload profiles, defined in Table~\ref{tab:workload-mixtures}, that span the three negotiable requirements at once and instantiate the operating regimes of Section~\ref{sec:motivation}. \textit{Lax-background} combines low coverage and sample with a long refresh interval for low-demand background analytics. \textit{Short-burst} combines a high sample fraction with a shorter refresh interval for event monitoring. \textit{Aggressive-incident} pushes coverage and sample high while holding freshness to a few seconds, the regime in which a stale snapshot would hide the forming jam from the emergency operator. \textit{Standard-operations} captures balanced day-to-day demand, and \textit{Cost-sensitive} narrows coverage and sample to respect a tight cost budget. The set covers both low-pressure and saturation regimes and defines a controlled operating envelope for comparing learned and non-learning policies rather than an exhaustive contract catalog.

\begin{table}[!htpb]
\centering
\begin{tabular}{llcc}
\toprule
Node & Role & vCPUs & Memory (GB) \\
\midrule
Large  & Orchestrator and worker & 16 & 32 \\
Medium & Worker                  & 2  & 4  \\
Small  & Worker                  & 1  & 2  \\
\bottomrule
\end{tabular}
\caption{Testbed nodes of the reference deployment.}
\label{tab:testbed-nodes}
\end{table}

\begin{table}[!htpb]
\centering
\begin{tabular}{lccc}
\toprule
Profile & Coverage & Sample & Freshness (s) \\
\midrule
Lax-background       & $[0.25, 0.45]$ & $[0.20, 0.45]$ & $[90, 180]$  \\
Short-burst          & $[0.60, 0.95]$ & $[0.60, 0.90]$ & $[10, 45]$   \\
Aggressive-incident  & $[0.75, 1.00]$ & $[0.70, 1.00]$ & $[5, 30]$    \\
Standard-operations  & $[0.45, 0.75]$ & $[0.35, 0.65]$ & $[30, 90]$   \\
Cost-sensitive       & $[0.33, 0.67]$ & $[0.25, 0.55]$ & $[60, 150]$  \\
\bottomrule
\end{tabular}
\caption{Canonical workload profiles.}
\label{tab:workload-mixtures}
\end{table}

% Paragraph 4: Campaign design and provenance
Each campaign cell couples one workload profile, one runtime, and one controller. Learned controllers train for 2048 decisions at the one-second cadence under the input-load schedule $\{2,16,64,4\}$, with 16 decisions per phase and 32 complete cycles. Their policies are serialized with a SHA-256 hash and a parameter fingerprint. Each artifact is then evaluated for 64 decisions with exploration and parameter updates disabled under the held-out schedule $\{3,12,48,6\}$. The \textit{best-fixed} reference is selected from the eight contract vertices and the joint midpoint using the tuning seeds $\{101,202,303\}$. Training uses $\{11,22,33,44,55\}$ and final evaluation uses $\{111,222,333,444,555\}$, so training, tuning, and evaluation have disjoint random streams.

% Paragraph 5: Campaign acceptance and accepted totals
A session enters the campaign index only when it completes its declared decisions and passes the provenance checks of Section~\ref{subsec:mdp}. Failed attempts are isolated and retained as excluded entries with an explicit reason. Per runtime, the accepted campaign comprises 75 training runs, three learners over five profiles and five seeds, 135 tuning runs, nine candidates over five profiles and three tuning seeds, and 150 frozen evaluations, six controllers over five profiles and five held-out seeds. The two runtimes together contribute 720 accepted sessions, and the 300 frozen evaluations score 19{,}200 decisions in total. The tuning and frozen-evaluation sessions reported here use the contract-feasible placement rule of Section~\ref{subsec:components}, with every learned policy artifact verified against its training SHA-256 hash and parameter fingerprint before loading.

% Paragraph 5: Metrics
The analysis rests on three measurements. The per-step reward of each evaluation is recomputed from the persisted reward components, so every aggregate can be audited offline. The paired delta compares a learned controller with each non-learning comparator on the same runtime, profile, and held-out seed. Each seed contributes the mean of its five profile-level deltas. The analysis reports the paired mean and median, a two-sided 95\,\% Student-t interval over the five seed means, the rank-biserial effect size, and win, tie, and loss counts. With only five paired seed means, the exact two-sided Wilcoxon signed-rank test cannot yield a p-value below $0.0625$, so the analysis emphasizes effect direction, effect size, seed win counts, and confidence intervals rather than confirmatory significance testing. In the live stage, contract and resource violations are recorded at the decision-epoch level for admitted requests, while queued-only requests are reported separately and do not count as frozen-policy evaluations. Requests admitted too close to the end of a trial to complete one frozen decision epoch are likewise reported separately as tail-censored, so a measurement boundary is never counted as a policy failure.

% Paragraph 6: Implementation and reproducibility
ARGOS is implemented in Python, with the orchestrator and worker-node interfaces exposed through FastAPI. The distributed profile uses the Neural Autonomic Transport System (NATS), a lightweight open-source messaging system, to exchange coordination events for capacity-aware leader election, while the local profile uses in-process coordination. The framework configuration, including the workload profiles of Table~\ref{tab:workload-mixtures} and the reward weights of Section~\ref{subsec:mdp}, is loaded from YAML at startup. The DQN and PPO controllers use the PyTorch backend, and every run records its active backend so that any fallback is excluded from the comparison. Decision, telemetry, lifecycle, and violation streams are stored as append-only JSON Lines (JSONL) files. Given a fixed set of accepted evidence packages, the analysis pipeline regenerates every aggregate table and figure deterministically, although the physical telemetry underneath is observed rather than assumed to reproduce bit for bit across executions. To respect space limitations, the paper reports the tables and figures most directly supporting the controlled comparison, resource-pressure analysis, and live validation. The complete result tables, additional figures, and scripts used to regenerate them are available in the public ARGOS artefact~\cite{MateosBravo2026ARGOS}.

\subsection{Controlled comparison}
\label{subsec:bench-results}

% Paragraph 7: Ranking question and headline
This subsection answers the model-ranking question. Table~\ref{tab:controlled-deltas} reports, for every learned controller, its paired reward-per-step advantage over each non-learning comparator on both runtimes. Each cell gives the mean delta, with positive values favoring the learned controller, the number of evaluation seeds that favor it out of five in parentheses, and the fraction of the \textit{best-fixed} reward that it recovers. The 95\,\% Student-t interval of every cell excludes zero except for Q-learning against the static midpoint and PPO against the threshold. Each seed contributes one profile-averaged delta, so the five evaluation seeds define the paired sample, and the sign is read as the learned controller minus the comparator. The picture is consistent across thread and process. DQN and PPO improve on the static midpoint in every seed but stay below the reactive threshold on average, Q-learning does not improve on the static midpoint, and all three learned controllers stay below the independently tuned \textit{best-fixed} reference.

\begin{table}[!htpb]
\centering
\begin{tabular}{llcccc}
\toprule
Runtime & Controller & vs.\ Static & vs.\ Threshold & vs.\ best-fixed & BF recovered \\
\midrule
\multirow{3}{*}{Thread}
 & DQN        & $+0.148$ (5/5) & $-0.079$ (0/5) & $-0.168$ (0/5) & $69\,\%$ \\
 & PPO        & $+0.161$ (5/5) & $-0.066$ (1/5) & $-0.154$ (0/5) & $72\,\%$ \\
 & Q-learning & $-0.024$ (1/5) & $-0.251$ (0/5) & $-0.339$ (0/5) & $38\,\%$ \\
\midrule
\multirow{3}{*}{Process}
 & DQN        & $+0.141$ (5/5) & $-0.093$ (1/5) & $-0.181$ (0/5) & $67\,\%$ \\
 & PPO        & $+0.170$ (5/5) & $-0.063$ (1/5) & $-0.152$ (0/5) & $72\,\%$ \\
 & Q-learning & $-0.038$ (2/5) & $-0.272$ (0/5) & $-0.360$ (0/5) & $34\,\%$ \\
\bottomrule
\end{tabular}
\caption{Paired reward advantage of each learned controller over each comparator.}
\label{tab:controlled-deltas}
\end{table}

\begin{figure}[!htpb]
  \centering
  \begin{subfigure}[t]{0.45\linewidth}
    \centering
    \includegraphics[width=\linewidth]{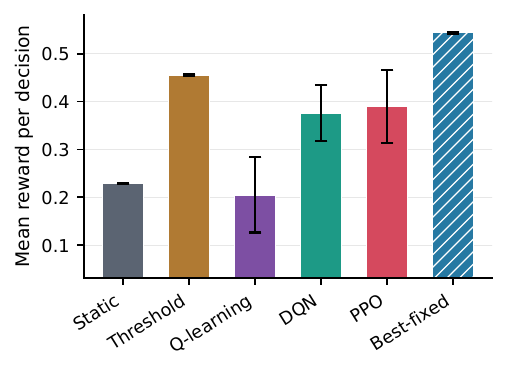}
    \caption{Thread runtime.}
    \label{fig:controlled-reward-thread}
  \end{subfigure}\hfill
  \begin{subfigure}[t]{0.45\linewidth}
    \centering
    \includegraphics[width=\linewidth]{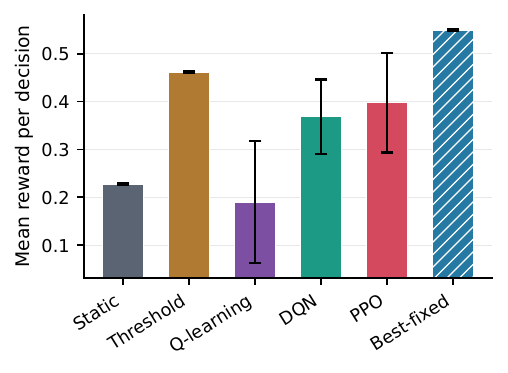}
    \caption{Process runtime.}
    \label{fig:controlled-reward-process}
  \end{subfigure}
  \caption{Mean frozen reward per decision for each controller.}
  \label{fig:controlled-reward}
\end{figure}

% Paragraph 8: Deep controllers against deployable baselines
DQN and PPO improve on the static midpoint on both runtimes, and the advantage holds in all five seeds of every cell. Figure~\ref{fig:controlled-reward} shows the same comparison as mean frozen reward with held-out-seed intervals. Averaged over the five profiles, DQN and PPO reach a reward per step of $0.376$ and $0.390$ on thread and $0.368$ and $0.398$ on process, against $0.229$ and $0.227$ for the static midpoint. The reactive threshold reaches $0.456$ and $0.461$, above both deep controllers on average. PPO holds a small mean-reward edge over DQN on both runtimes, although the two deep controllers are better described as a tier than as a strict order.

\begin{figure}[!htpb]
  \centering
  \begin{subfigure}[b]{0.8\linewidth}
    \centering
    \includegraphics[width=\linewidth]{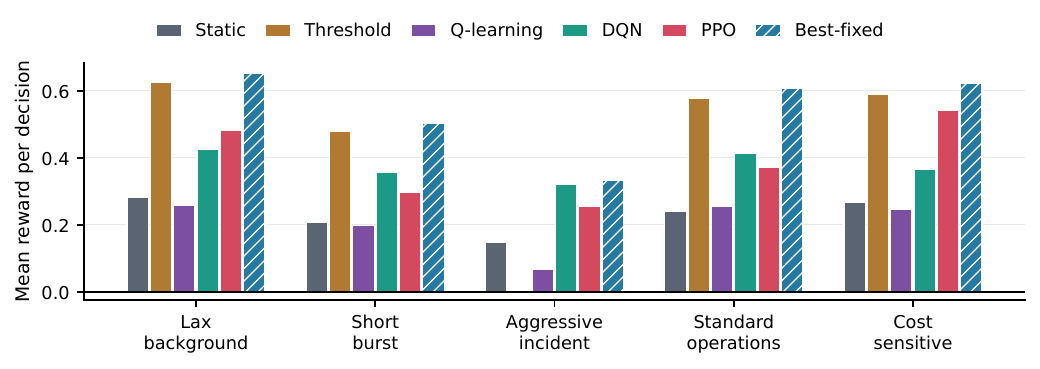}
    \caption{Thread runtime.}
    \label{fig:controlled-profiles-thread}
  \end{subfigure}\\[0.4em]
  \begin{subfigure}[b]{0.8\linewidth}
    \centering
    \includegraphics[width=\linewidth]{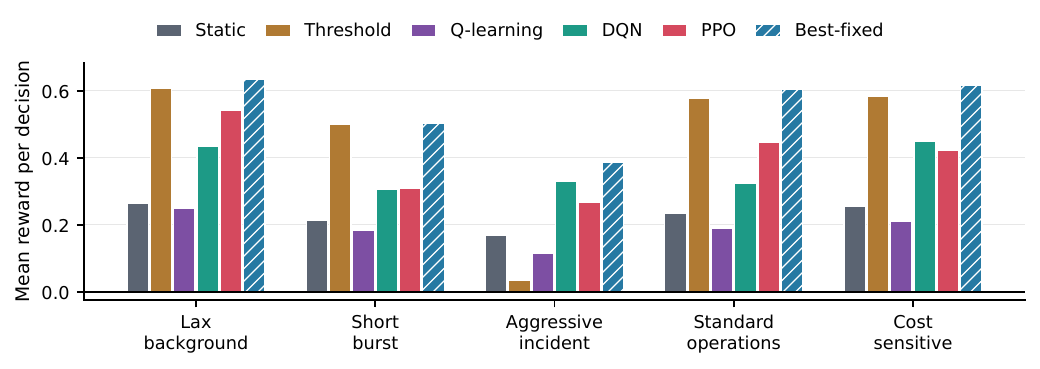}
    \caption{Process runtime.}
    \label{fig:controlled-profiles-process}
  \end{subfigure}
  \caption{Mean frozen reward per decision by workload profile and controller.}
  \label{fig:controlled-profiles}
\end{figure}

% Paragraph 9: Profile-level behavior
Figure~\ref{fig:controlled-profiles} decomposes the comparison across the five workload profiles. The deep controllers stay above the static midpoint in every profile on both runtimes, while neither dominates the other everywhere. The reactive threshold behaves differently according to resource pressure. In lax-background, short-burst, standard-operations, and cost-sensitive, it remains near the highest-quality feasible configuration and is competitive with the deep controllers. Under aggressive-incident, sustained pressure activates its downward quality cycle, reducing its reward to $0.006$ on thread and $0.035$ on process, while DQN and PPO reach $0.255$ to $0.330$. The \textit{best-fixed} reference remains strongest in every profile.

% Paragraph 10: Tabular controller does not clear the baselines
The tabular Q-learning controller is not a deployable improvement. It fails to improve on the static midpoint on both runtimes, with paired deltas of $-0.024$ and $-0.038$ reward per step, only one and two favorable seeds, and intervals that include zero. It also stays below the reactive threshold on both runtimes, by $0.251$ and $0.272$ reward per step. This result is consistent with the difficulty of covering the thirteen-dimensional discrete state of Section~\ref{subsec:mdp} within the available training budget. Q-learning therefore serves as a lightweight tabular reference rather than a competitive deployment candidate.

% Paragraph 11: Best-fixed as an independently tuned reference
Every learned controller stays below the \textit{best-fixed} reference, by $0.152$ to $0.181$ reward per step for the two deep controllers and by roughly $0.35$ for Q-learning. This gap measures headroom against a comparator that is tuned per workload profile in advance and then held fixed throughout held-out evaluation. That the deep controllers recover $67$ to $72\,\%$ of its reward while making online per-request control decisions is the main result of the controlled stage.

\subsection{Resource-pressure analysis}
\label{subsec:resources}

% Paragraph 12: How controllers load the constrained node
The controlled campaign also records how each controller loads the constrained node, which exposes the quality-against-pressure trade that motivates adaptive control. Spatial fidelity is $1-\frac{1}{2}\sum_i |p_i-\hat p_i|$, where $p_i$ and $\hat p_i$ are normalized cell frequencies in the sampled heatmap and its full-sample reference for the same trajectory partition. Table~\ref{tab:resource-pressure} pairs this measure with the mean service duty driven on the assigned nodes, the fraction of each refresh interval spent computing, and the 95th-percentile cluster CPU utilization. Every entry averages the five profiles and five seeds of its runtime. The deep controllers reach a spatial fidelity of $0.935$ to $0.945$, above the $0.926$ of the static midpoint and within $0.025$ of the $0.958$ to $0.959$ \textit{best-fixed} reference. On thread, PPO reaches $0.942$ fidelity at $9.2\,\%$ duty, compared with $18.6\,\%$ for the \textit{best-fixed} reference.

% Paragraph 13: The trade the controllers make
The extremes of Table~\ref{tab:resource-pressure} show the trade-off. The static midpoint keeps the lowest service duty but also the lowest fidelity. The reactive threshold reaches a fidelity of $0.948$ to $0.949$ at $7$ to $8\,\%$ duty, consistent with its strong performance in profiles without sustained pressure. The \textit{best-fixed} reference reaches the highest fidelity and service duty. Among the deep controllers, PPO uses the lowest duty on both runtimes, while DQN reaches a similar fidelity at higher duty. Cluster CPU p95 remains between $50$ and $52\,\%$ on thread and between $27$ and $52\,\%$ on process.

\begin{table}[!htpb]
\centering
\begin{tabular}{lcccccc}
\toprule
 & \multicolumn{3}{c}{Thread} & \multicolumn{3}{c}{Process} \\
\cmidrule(lr){2-4}\cmidrule(lr){5-7}
Controller & Fidelity & Duty & CPU p95 & Fidelity & Duty & CPU p95 \\
\midrule
best-fixed & $0.958$ & $18.6\,\%$ & $51\,\%$ & $0.959$ & $15.0\,\%$ & $42\,\%$ \\
DQN        & $0.945$ & $11.2\,\%$ & $51\,\%$ & $0.935$ & $12.8\,\%$ & $50\,\%$ \\
PPO        & $0.942$ & $9.2\,\%$  & $51\,\%$ & $0.944$ & $8.9\,\%$  & $35\,\%$ \\
Q-learning & $0.925$ & $7.6\,\%$  & $52\,\%$ & $0.927$ & $6.1\,\%$  & $52\,\%$ \\
Threshold  & $0.949$ & $8.0\,\%$  & $51\,\%$ & $0.948$ & $7.0\,\%$  & $43\,\%$ \\
Static     & $0.926$ & $5.9\,\%$  & $50\,\%$ & $0.926$ & $4.3\,\%$  & $27\,\%$ \\
\bottomrule
\end{tabular}
\caption{Spatial fidelity, service duty, and CPU utilization per controller; p95 denotes the 95th percentile.}
\label{tab:resource-pressure}
\end{table}

\subsection{Live-trial validation}
\label{subsec:live}

% Paragraph 14: Purpose of the live stage
The live stage shifts the question from model ranking to deployability under load. Whereas the controlled stage drives one request through a fixed schedule, the live stage submits requests that arrive and depart over time under two arrival regimes: a realistic regime that keeps the cluster below saturation and a concurrency regime that pushes it into overload. Admitted requests that breach any contract or resource bound are reported as service-level objective violations. This separates operational safety, whether the deployed controller keeps the analytics inside the client ranges while the cluster is contended, from the reward ranking of the controlled stage.

% Paragraph 15: Live design
The live campaign pairs the static midpoint with each deep controller, DQN and PPO, on the thread runtime. Tabular Q-learning is not carried into this stage because it does not improve on the static midpoint in the controlled comparison, so it is not a deployment candidate, and the live stage evaluates the controllers a deployment would actually consider. The controlled stage already covers both runtimes, so the live stage fixes one and spends its budget on paired repetitions instead.

% Paragraph 16: Live campaign protocol and accepted totals
Each arrival regime is repeated over five live seeds, and each seed drives one static, one DQN, and one PPO trial of 30 minutes with an identical arrival trace, so every comparison is paired at the trace level. Each learned trial loads the frozen artifact trained under the matching training seed and applies it to every admitted request without exploration or parameter updates, while admission, queueing, and placement remain active in every variant. The 30 accepted trials cover 62{,}976 decision epochs and 393 admitted requests that completed at least one frozen decision epoch. Two further requests were tail-censored, and no trial attempt was excluded.

\begin{figure}[!htpb]
  \centering
  \begin{subfigure}[t]{0.44\linewidth}
    \centering
    \includegraphics[width=\linewidth]{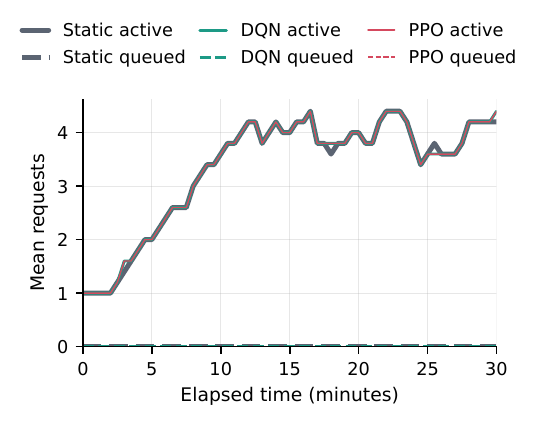}
    \caption{Realistic regime.}
    \label{fig:live-load-realistic}
  \end{subfigure}\hfill
  \begin{subfigure}[t]{0.44\linewidth}
    \centering
    \includegraphics[width=\linewidth]{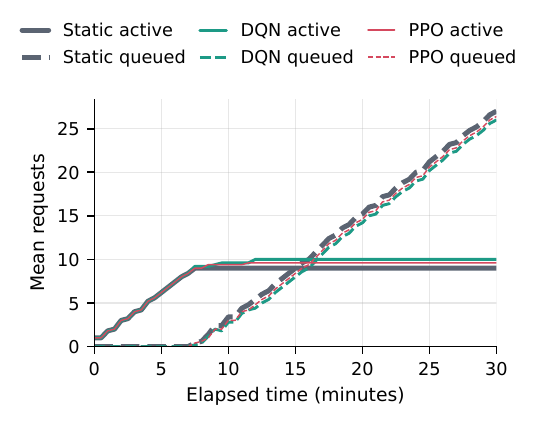}
    \caption{Concurrency regime.}
    \label{fig:live-load-concurrency}
  \end{subfigure}
  \caption{Mean active and queued requests over time in the live stage.}
  \label{fig:live-load}
\end{figure}

% Paragraph 16: Arrival regimes and admission
Figure~\ref{fig:live-load} shows the offered load, averaged over the five paired seeds of each variant. Where the three curves coincide, the thinner DQN and PPO lines are drawn over the wider static line. The realistic regime (Figure~\ref{fig:live-load-realistic}) stays below saturation: every submitted request is admitted on arrival, the active set plateaus around four requests, and the queue stays empty in every variant. The concurrency regime (Figure~\ref{fig:live-load-concurrency}) overloads the cluster by design: of the 180 requests submitted per variant, about 46 are admitted on arrival while about 134 wait in the pending queue, which grows toward 27 requests as the active set saturates at nine to ten requests. Admission behaves almost identically under all three controllers, with overlapping active and queued trajectories, so the arrival regime rather than the quality controller determines what enters the cluster, and the paired comparison isolates what each controller does with the admitted work.

\begin{figure}[!htpb]
  \centering
  \begin{subfigure}[t]{0.44\linewidth}
    \centering
    \includegraphics[width=\linewidth]{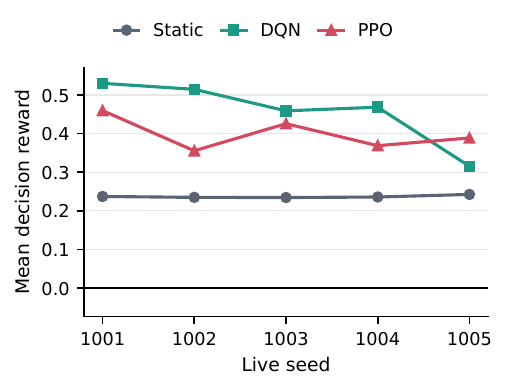}
    \caption{Realistic regime.}
    \label{fig:live-reward-realistic}
  \end{subfigure}\hfill
  \begin{subfigure}[t]{0.44\linewidth}
    \centering
    \includegraphics[width=\linewidth]{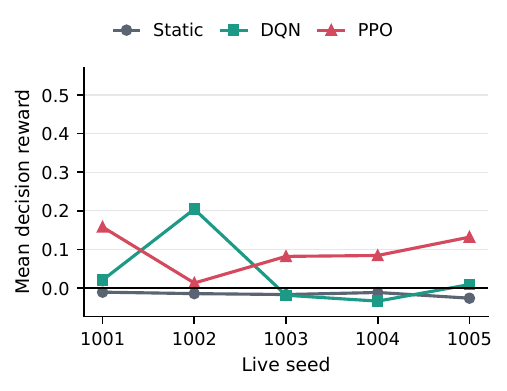}
    \caption{Concurrency regime.}
    \label{fig:live-reward-concurrency}
  \end{subfigure}
  \caption{Paired mean decision reward per live seed.}
  \label{fig:live-reward}
\end{figure}

% Paragraph 17: Paired live outcome
Figure~\ref{fig:live-reward} reports the paired outcome. Under realistic arrivals, DQN improves mean decision reward over the static midpoint by $+0.221$ with a 95\,\% interval of $[0.112,0.330]$, and PPO by $+0.163$ with an interval of $[0.110,0.216]$, with improvements in all five pairs. Under overload, PPO improves by $+0.110$ in all five pairs with an interval of $[0.039,0.180]$, whereas DQN improves by $+0.052$ in three of five pairs with an interval of $[-0.067,0.172]$. Spatial fidelity shows the same pattern: PPO improves in all ten pairs, while DQN is consistent under realistic arrivals but more variable under overload.

% Paragraph 18: Contract exposure
Contract exposure completes the safety reading. Table~\ref{tab:live-exposure} reports it over the five paired seeds, with admitted requests that completed at least one frozen decision epoch, the mean fraction of decision epochs carrying any violation, and violation counts by dimension. Across the 62{,}976 decision epochs of the 30 trials, the only recorded violation events are 21 freshness events, with zero coverage, sample, CPU, or memory events.

\begin{table}[!htpb]
\centering
\begin{tabular}{llccccc}
\toprule
Regime & Variant & Evaluated &
\shortstack{Violated\\epochs} &
\shortstack{Coverage\\viol.} &
\shortstack{Freshness\\viol.} &
\shortstack{Sample/CPU/mem.\\viol.} \\
\midrule
\multirow{3}{*}{Realistic}
 & Static & $50$ & $0.00\,\%$ & $0$ & $0$  & $0$ \\
 & DQN    & $50$ & $0.00\,\%$ & $0$ & $0$  & $0$ \\
 & PPO    & $50$ & $0.00\,\%$ & $0$ & $0$  & $0$ \\
\midrule
\multirow{3}{*}{Concurrency}
 & Static & $80$ & $0.00\,\%$ & $0$ & $0$  & $0$ \\
 & DQN    & $82$ & $0.02\,\%$ & $0$ & $3$  & $0$ \\
 & PPO    & $81$ & $0.12\,\%$ & $0$ & $18$ & $0$ \\
\bottomrule
\end{tabular}

\vspace{1mm}
{\footnotesize \textit{Note:} viol. = violation; mem. = memory.}

\caption{Live contract exposure per regime and variant.}
\label{tab:live-exposure}
\end{table}

% Paragraph 19: Coverage safety under discrete placement
The fraction of violated decision epochs is zero in every realistic trial and at most $0.12\,\%$ under overload. The live trials therefore support operational feasibility on the evaluated testbed rather than a general guarantee of contract compliance.
\section{Discussion}
\label{sec:discussion}

This section interprets the experimental evidence in relation to the objectives established in the preceding sections and separates the conclusions supported by the results from the limitations that bound their scope. Section~\ref{subsec:discussion-results} discusses the controlled and live-stage findings and their operational implications, while Section~\ref{subsec:limitations} examines the internal, external, scalability, and construct limitations of the study together with the scope of its reproducibility claims.

\subsection{Results discussion}
\label{subsec:discussion-results}

% Paragraph 1: Main controlled claim
The controlled evidence shows that DQN and PPO improve on the static midpoint in every seed of both runtimes. The reactive threshold is competitive when sustained resource pressure is absent, but the deep controllers outperform it by a wide margin under the aggressive-incident profile, where pressure persists. Q-learning does not improve on the static midpoint and remains below the threshold. PPO holds a small mean-reward edge over DQN, although the two deep controllers are better described as a tier than as a strict order.

% Paragraph 2: Independently tuned reference and resource pressure
All three learned controllers stay below the \textit{best-fixed} reference, and that gap is read as headroom against a strong static comparator rather than against an oracle. A deployment can use the selected point when its workload profile is known and stable, but that fixed configuration cannot react to changing node pressure during a request. The deep controllers recover $67$ to $72\,\%$ of its reward, trailing it by about $0.15$ to $0.18$ reward per step, while retaining online per-request control. The resource-pressure analysis adds an operational reading: the deep controllers lift spatial fidelity above the static midpoint without requiring the highest service pressure. PPO holds the lowest service duty of the deep controllers on both runtimes, and neither deep controller exceeds the duty of the \textit{best-fixed} reference. These results show that adaptive quality control can use available node headroom to improve analytics quality without requiring the highest service duty in every setting, which is the behavior motivated in Section~\ref{sec:motivation}.

% Paragraph 3: Live-stage reading
The live stage of Section~\ref{subsec:live} extends this reading to multi-tenant operation. Paired on identical arrival traces, both frozen deep controllers improve mean decision reward and delivered fidelity over the static midpoint. Under overload, both retain positive mean reward while the static midpoint has negative mean reward. They differ in consistency: PPO improves in every overload pair with an interval that excludes zero, whereas DQN improves in three of five pairs with an interval that includes zero. With contract-feasible placement, no coverage, sample, CPU, or memory violations are recorded, while the 21 freshness events keep the violated-epoch fraction at or below $0.12\,\%$. The live evidence therefore supports deployability of the frozen deep controllers rather than providing a second model ranking, preserving the separation intended by the two-stage protocol.

% Paragraph 4: Practical implications
For deployments resembling the evaluated setting, either deep controller is a reasonable candidate when reward per request matters. PPO provides the highest controlled mean reward among the learned controllers and the lowest mean service duty of the deep tier, while DQN posts the larger live gain under realistic arrivals. PPO also shows the more consistent overload behavior in the live trials, while DQN's advantage varies more across seeds. Q-learning is better treated as a tabular reference than as a deployment candidate because it does not improve on the static midpoint on either runtime. The static midpoint remains a deterministic fallback when operational simplicity outweighs the adaptive gain. The \textit{best-fixed} reference suits a known and stable workload profile when offline tuning is acceptable, but it remains a comparator rather than a general adaptive solution because its operating point cannot respond to changing resource pressure.

\subsection{Limitations}
\label{subsec:limitations}

% Paragraph 5: Internal and external validity
Several factors bound how broadly the contribution transfers. On the internal side, the paired protocol shares the seed schedule, orchestration loop, persistence path, and comparison scripts across controllers, which reduces implementation-level confounds. The learned policies were trained before the contract-feasible placement rule was introduced and were evaluated frozen under the corrected mapping. Retraining under the revised placement rule is left for future work. The main statistical limitation is the number of paired evaluation seeds. Each comparison rests on five seed means, so the exact two-sided Wilcoxon signed-rank test cannot yield a p-value below $0.0625$. The Student-t intervals assume approximately normal seed-level mean differences, an assumption that is difficult to assess with only five seeds. The analysis therefore relies on effect direction, effect size, seed win counts, and confidence intervals rather than on a confirmatory significance threshold. On the external side, the workload consists of one urban-mobility heatmap service over simulated Seville movement trajectories, and the five workload profiles cover selected operating regimes without exhausting the contract space. The testbed uses three virtual machines in one cloud network and includes no mist-tier devices, emulated wide-area network conditions, or direct energy measurements. The controlled comparison covers both thread and process runtimes, whereas the live stage uses the thread runtime, so its deployability evidence does not cover every runtime or application setting.

% Paragraph 6: Scalability
Scalability beyond the evaluated testbed remains an open question. The controlled and live stages use three heterogeneous nodes, which is sufficient to exercise heterogeneous placement, resource contention, admission, and per-request adaptation, but does not characterize how ARGOS behaves as the number of nodes and concurrent requests grows. Each learned policy is trained with one active request, whereas the live concurrency regime holds nine to ten admitted requests at its plateau and at most eleven. The live stage therefore evaluates transfer to multi-tenant contention rather than training under it. Larger deployments may change placement granularity, telemetry and coordination overhead, and the interaction between admission and per-request control. The distributed leader-failover mechanism is also outside the present evaluation, which keeps one active orchestrator to isolate the quality-control behavior. Evaluating these aspects on larger Continuum topologies is therefore an important direction for future work.

% Paragraph 7: Construct validity
On the construct side, the reward function uses fixed coefficients of $1.0$ for quality, $1.2$ for resource pressure, and $0.5$ for cost, with resource and cost normalized into a shared penalty budget. These coefficients are fixed design parameters rather than the result of per-deployment optimization. The measurement conventions introduce additional limits. Quality and cost reflect the selected configuration, while resource telemetry comes from the preceding poll and may not yet reflect that configuration's resource effects. This limits causal interpretation of an individual action's immediate resource effect. Spatial fidelity measures sampling distortion within the evaluated partitions, not city-wide coverage or result freshness. The 20-epoch violation count is also a bounded-history summary. It captures recent violation frequency but not the exact order or age of individual events, so the decision model approximates Markov observability from the telemetry available to the controller. Finally, the action space is discrete, which improves auditability but leaves continuous control unexplored.

% Paragraph 8: Evaluation realism and reproducibility
These limitations should be considered together with the evaluation protocol. ARGOS is exercised on a deployed heterogeneous cluster using the resource telemetry produced under actual contention. Reproducibility studies have shown that single-seed evaluations, uncontrolled hyperparameter choices, and insufficient separation between training and evaluation can make deep reinforcement-learning results fragile~\cite{Henderson2018DeepRLMatters,Colas2018HowManySeeds}. The orchestration studies discussed in Section~\ref{sec:related} vary in how fully they address these concerns. Evaluating the controllers under capacity pressure and scoring frozen policies with held-out seeds and checked policy fingerprints therefore strengthens the deployment relevance and auditability of the reported comparison~\cite{Henderson2018DeepRLMatters,Pineau2021Reproducibility}.
\section{Conclusions and future work}
\label{sec:conclusions}

% Paragraph 1: Problem and ARGOS response
Computing Continuum analytics need more than resource scaling when heterogeneous nodes become capacity-limited. Static quality policies cannot react to cluster pressure, while placement or replica controllers do not act on the analytics requirements that define each contract. ARGOS addresses this problem by operationalizing multidimensional elasticity as a per-request MDP that adjusts coverage, sample, and freshness within the ranges accepted by each client and combines those decisions with capacity-aware admission. Every controller, learned or not, shares the same orchestration loop and persistence path. Each learned policy is frozen and fingerprinted before evaluation, so the reported gains are derived from auditable artifacts evaluated over multiple held-out seeds.

% Paragraph 2: Evidence and claim boundaries
Across two runtimes, five workload profiles, and five held-out evaluation seeds, DQN and PPO improve on the static midpoint in every seed and outperform the reactive threshold under the sustained pressure of the aggressive-incident profile. They recover $67$ to $72\,\%$ of the reward of the independently tuned \textit{best-fixed} reference at a lower mean service duty on both runtimes. Tabular Q-learning does not improve on the static midpoint on either runtime, so the evidence supports the deep controllers rather than reinforcement learning in general. With only five paired seed means, the analysis emphasizes effect direction, effect size, seed win counts, and confidence intervals rather than confirmatory significance testing. The live stage of Section~\ref{subsec:live} complements the controlled comparison under time-varying arrivals. Both frozen deep controllers improve mean reward and delivered fidelity over the static midpoint, while PPO improves in every paired trial of both arrival regimes. With contract-feasible placement, no coverage, sample, CPU, or memory violations are recorded, and freshness violations are rare. The controlled and live stages therefore support distinct claims about comparative performance and deployability.

% Paragraph 3: Future work
Future work should extend ARGOS in three directions. First, continuous-action approaches should be evaluated to determine whether finer control can improve on the fixed coverage, sample, and freshness step sizes used here. Second, evaluation should include more seeds and additional deployment configurations to increase statistical power, particularly for differences within the deep-controller tier and under live traffic. Third, larger Continuum topologies and additional analytics services should be studied to determine how multidimensional quality control interacts with placement and whether the observed behavior generalizes beyond the three-node urban-mobility setting.

\section*{Acknowledgment}

This work has been partially funded by the grant PTQ2024-013825 funded by MICIU/AEI/10.13039/50 1100011033, and co-financed 85\% by the European Union, the European Regional Development Fund and by the Department of Education, Science and Professional Training of the Regional Government of Extremadura. Managing Authority: Ministry of Finance. Grant File Number: GR24099. This work has also been developed within the framework of the POCTEP project ``Gestión preventiva del patrimonio construido para su protección frente al cambio climático a través de la promoción del turismo sostenible e inclusivo'' (HEPRESTONE).

%, by the Department of Education, Science and Professional Training of the Government of Extremadura (GR24099), and by the European Regional Development Fund.

\bibliographystyle{elsarticle-num}
\bibliography{bibliography}

@software{MateosBravo2026ARGOS,
  author       = {Javier Mateos-Bravo and Sergio Laso and Juan Luis Herrera and Ilir Murturi and Pantelis Frangoudis and Schahram Dustdar},
  title        = {ARGOS: Reinforcement Learning-Driven Multidimensional Elasticity for Service Orchestration in the Computing Continuum},
  year         = {2026},
  version      = {1.0.0},
  publisher    = {Zenodo},
  doi          = {10.5281/zenodo.22933898},
  url          = {https://doi.org/10.5281/zenodo.22933898}
}

@inproceedings{Murturi2025ElasticityIoT,
  author    = {Sergio Laso and Ilir Murturi and Pantelis Frangoudis and Juan Luis Herrera and Juan M. Murillo and Schahram Dustdar},
  title     = {A Multidimensional Elasticity Framework for Adaptive Data Analytics Management in the Computing Continuum},
  booktitle = {Proceedings of the 2025 IEEE International Conference on Communications (ICC)},
  pages     = {1133--1138},
  publisher = {IEEE},
  year      = {2025},
  doi       = {10.1109/ICC52391.2025.11161512}
}

@article{Dustdar2011ElasticProcesses,
  author  = {Schahram Dustdar and Yike Guo and Benjamin Satzger and Hong-Linh Truong},
  title   = {Principles of Elastic Processes},
  journal = {IEEE Internet Computing},
  volume  = {15},
  number  = {5},
  pages   = {66--71},
  year    = {2011},
  doi     = {10.1109/MIC.2011.121}
}

@article{Murturi2022DECENT,
  author  = {Ilir Murturi and Schahram Dustdar},
  title   = {{DECENT}: A Decentralized Configurator for Controlling Elasticity in Dynamic Edge Networks},
  journal = {ACM Transactions on Internet Technology},
  volume  = {22},
  number  = {3},
  pages   = {78:1--78:21},
  year    = {2022},
  doi     = {10.1145/3530692}
}

@article{Casamayor2023EdgeIntelligence,
  author  = {V{\'i}ctor Casamayor Pujol and Praveen Kumar Donta and Andrea Morichetta and Ilir Murturi and Schahram Dustdar},
  title   = {Edge Intelligence: Research Opportunities for Distributed Computing Continuum Systems},
  journal = {IEEE Internet Computing},
  volume  = {27},
  number  = {4},
  pages   = {53--74},
  year    = {2023},
  doi     = {10.1109/MIC.2023.3284693}
}

@article{LoridoBotran2014AutoscalingReview,
  author  = {Tania Lorido-Botran and Jos{\'e} Miguel-Alonso and Jose Antonio Lozano},
  title   = {A Review of Auto-scaling Techniques for Elastic Applications in Cloud Environments},
  journal = {Journal of Grid Computing},
  volume  = {12},
  number  = {4},
  pages   = {559--592},
  year    = {2014},
  doi     = {10.1007/s10723-014-9314-7}
}

@article{Casalicchio2019Autoscaling,
  author  = {Emiliano Casalicchio},
  title   = {A Study on Performance Measures for Auto-Scaling {CPU}-Intensive Containerized Applications},
  journal = {Cluster Computing},
  volume  = {22},
  number  = {3},
  pages   = {995--1006},
  year    = {2019},
  doi     = {10.1007/s10586-018-02890-1}
}

@inproceedings{Khaleq2021QoSAwareAutoscaling,
  author    = {Abeer Abdel Khaleq and Ilkyeun Ra},
  title     = {Development of {QoS}-Aware Agents with Reinforcement Learning for Autoscaling of Microservices on the Cloud},
  booktitle = {Proceedings of the 2021 IEEE International Conference on Autonomic Computing and Self-Organizing Systems Companion (ACSOS-C)},
  pages     = {13--19},
  year      = {2021},
  doi       = {10.1109/ACSOS-C52956.2021.00025}
}

@inproceedings{Trihinas2015AdaM,
  author    = {Demetris Trihinas and George Pallis and Marios D. Dikaiakos},
  title     = {{AdaM}: An Adaptive Monitoring Framework for Sampling and Filtering on {IoT} Devices},
  booktitle = {Proceedings of the 2015 IEEE International Conference on Big Data (Big Data)},
  pages     = {717--726},
  year      = {2015},
  doi       = {10.1109/BigData.2015.7363816}
}

@book{SuttonBarto2018RL,
  author    = {Richard S. Sutton and Andrew G. Barto},
  title     = {Reinforcement Learning: An Introduction},
  edition   = {2nd},
  publisher = {MIT Press},
  address   = {Cambridge, MA},
  year      = {2018}
}

@article{Watkins1992QLearning,
  author  = {Christopher J. C. H. Watkins and Peter Dayan},
  title   = {Technical Note: {Q}-learning},
  journal = {Machine Learning},
  volume  = {8},
  number  = {3--4},
  pages   = {279--292},
  year    = {1992},
  doi     = {10.1007/BF00992698}
}

@inproceedings{Tesauro2006Hybrid,
  author    = {Gerald Tesauro and Nicholas K. Jong and Rajarshi Das and Mohamed N. Bennani},
  title     = {A Hybrid Reinforcement Learning Approach to Autonomic Resource Allocation},
  booktitle = {Proceedings of the 2006 IEEE International Conference on Autonomic Computing (ICAC)},
  pages     = {65--73},
  year      = {2006},
  doi       = {10.1109/ICAC.2006.1662383}
}

@inproceedings{Mao2016DeepRM,
  author    = {Hongzi Mao and Mohammad Alizadeh and Ishai Menache and Srikanth Kandula},
  title     = {Resource Management with Deep Reinforcement Learning},
  booktitle = {Proceedings of the 15th ACM Workshop on Hot Topics in Networks (HotNets)},
  pages     = {50--56},
  year      = {2016},
  doi       = {10.1145/3005745.3005750}
}

@inproceedings{Mao2019Decima,
  author    = {Hongzi Mao and Malte Schwarzkopf and Shaileshh Bojja Venkatakrishnan and Zili Meng and Mohammad Alizadeh},
  title     = {Learning Scheduling Algorithms for Data Processing Clusters},
  booktitle = {Proceedings of the ACM Special Interest Group on Data Communication (SIGCOMM)},
  pages     = {270--288},
  year      = {2019},
  doi       = {10.1145/3341302.3342080}
}

@inproceedings{Rossi2019HorizVertScaling,
  author    = {Fabiana Rossi and Matteo Nardelli and Valeria Cardellini},
  title     = {Horizontal and Vertical Scaling of Container-based Applications Using Reinforcement Learning},
  booktitle = {Proceedings of the 2019 IEEE 12th International Conference on Cloud Computing (CLOUD)},
  pages     = {329--338},
  year      = {2019},
  doi       = {10.1109/CLOUD.2019.00061}
}

@article{Gari2021QLearningAutoscaling,
  author  = {Yisel Gar{\'i} and David A. Monge and Cristian Mateos},
  title   = {A {Q}-learning Approach for the Autoscaling of Scientific Workflows in the Cloud},
  journal = {Future Generation Computer Systems},
  volume  = {127},
  pages   = {168--180},
  year    = {2022},
  doi     = {10.1016/j.future.2021.09.007}
}

@article{Goudarzi2024DistributedDRL,
  author  = {Mohammad Goudarzi and Marimuthu Palaniswami and Rajkumar Buyya},
  title   = {A Distributed Deep Reinforcement Learning Technique for Application Placement in Edge and Fog Computing Environments},
  journal = {IEEE Transactions on Mobile Computing},
  volume  = {22},
  number  = {5},
  pages   = {2491--2505},
  year    = {2023},
  doi     = {10.1109/TMC.2021.3123165}
}

@article{laso2025energy,
  title={Energy consumption and workload prediction for edge nodes in the Computing Continuum},
  author={Laso, Sergio and Rodriguez, Pablo and Herrera, Juan Luis and Berrocal, Javier and Murillo, Juan M},
  journal={Sustainable Computing: Informatics and Systems},
  volume={46},
  pages={101088},
  year={2025},
  publisher={Elsevier}
}

@article{WangGoudarzi2024EdgeFogScheduling,
  author  = {Zhiyu Wang and Mohammad Goudarzi and Mingming Gong and Rajkumar Buyya},
  title   = {Deep Reinforcement Learning-Based Scheduling for Optimizing System Load and Response Time in Edge and Fog Computing Environments},
  journal = {Future Generation Computer Systems},
  volume  = {152},
  pages   = {55--69},
  year    = {2024},
  doi     = {10.1016/j.future.2023.10.012}
}

@article{Mnih2015DQN,
  author  = {Volodymyr Mnih and Koray Kavukcuoglu and David Silver and Andrei A. Rusu and Joel Veness and Marc G. Bellemare and Alex Graves and Martin Riedmiller and Andreas K. Fidjeland and Georg Ostrovski and Stig Petersen and Charles Beattie and Amir Sadik and Ioannis Antonoglou and Helen King and Dharshan Kumaran and Daan Wierstra and Shane Legg and Demis Hassabis},
  title   = {Human-Level Control Through Deep Reinforcement Learning},
  journal = {Nature},
  volume  = {518},
  number  = {7540},
  pages   = {529--533},
  year    = {2015},
  doi     = {10.1038/nature14236}
}

@misc{Schulman2017PPO,
  author       = {John Schulman and Filip Wolski and Prafulla Dhariwal and Alec Radford and Oleg Klimov},
  title        = {Proximal Policy Optimization Algorithms},
  year         = {2017},
  eprint       = {1707.06347},
  archivePrefix= {arXiv},
  primaryClass = {cs.LG},
  doi          = {10.48550/arXiv.1707.06347}
}

@incollection{Bonomi2014FogPlatform,
  author    = {Flavio Bonomi and Rodolfo Milito and Preethi Natarajan and Jiang Zhu},
  title     = {Fog Computing: A Platform for Internet of Things and Analytics},
  booktitle = {Big Data and Internet of Things: A Roadmap for Smart Environments},
  editor    = {Nik Bessis and Ciprian Dobre},
  series    = {Studies in Computational Intelligence},
  volume    = {546},
  pages     = {169--186},
  publisher = {Springer},
  year      = {2014},
  doi       = {10.1007/978-3-319-05029-4_7}
}

@article{Satyanarayanan2017EdgeEmergence,
  author  = {Mahadev Satyanarayanan},
  title   = {The Emergence of Edge Computing},
  journal = {Computer},
  volume  = {50},
  number  = {1},
  pages   = {30--39},
  year    = {2017},
  doi     = {10.1109/MC.2017.9}
}

@article{Rossi2023DynamicThresholds,
  author  = {Fabiana Rossi and Valeria Cardellini and Francesco Lo Presti and Matteo Nardelli},
  title   = {Dynamic Multi-Metric Thresholds for Scaling Applications Using Reinforcement Learning},
  journal = {IEEE Transactions on Cloud Computing},
  volume  = {11},
  number  = {2},
  pages   = {1807--1821},
  year    = {2023},
  doi     = {10.1109/TCC.2022.3163357}
}

@article{Mampage2023ServerlessRL,
  author  = {Anupama Mampage and Shanika Karunasekera and Rajkumar Buyya},
  title   = {Deep Reinforcement Learning for Application Scheduling in Resource-Constrained, Multi-Tenant Serverless Computing Environments},
  journal = {Future Generation Computer Systems},
  volume  = {143},
  pages   = {277--292},
  year    = {2023},
  doi     = {10.1016/j.future.2023.02.006}
}

@article{GariRLAutoscalingSurvey2021,
  author  = {Yisel Gar{\'i} and David A. Monge and Elina Pacini and Cristian Mateos and Carlos Garc{\'i}a Garino},
  title   = {Reinforcement Learning-Based Application Autoscaling in the Cloud: A Survey},
  journal = {Engineering Applications of Artificial Intelligence},
  volume  = {102},
  pages   = {104288},
  year    = {2021},
  doi     = {10.1016/j.engappai.2021.104288}
}

@inproceedings{Henderson2018DeepRLMatters,
  author    = {Peter Henderson and Riashat Islam and Philip Bachman and Joelle Pineau and Doina Precup and David Meger},
  title     = {Deep Reinforcement Learning That Matters},
  booktitle = {Proceedings of the Thirty-Second AAAI Conference on Artificial Intelligence (AAAI)},
  pages     = {3207--3214},
  year      = {2018},
  doi       = {10.1609/aaai.v32i1.11694}
}

@misc{Colas2018HowManySeeds,
  author       = {C{\'e}dric Colas and Olivier Sigaud and Pierre-Yves Oudeyer},
  title        = {How Many Random Seeds? Statistical Power Analysis in Deep Reinforcement Learning Experiments},
  year         = {2018},
  eprint       = {1806.08295},
  archivePrefix= {arXiv},
  primaryClass = {cs.LG},
  doi          = {10.48550/arXiv.1806.08295}
}

@inproceedings{Agarwal2021Precipice,
  author    = {Rishabh Agarwal and Max Schwarzer and Pablo Samuel Castro and Aaron C. Courville and Marc G. Bellemare},
  title     = {Deep Reinforcement Learning at the Edge of the Statistical Precipice},
  booktitle = {Advances in Neural Information Processing Systems 34 (NeurIPS)},
  pages     = {29304--29320},
  year      = {2021},
  url       = {https://proceedings.neurips.cc/paper/2021/hash/f514cec81cb148559cf475e7426eed5e-Abstract.html}
}

@article{Pineau2021Reproducibility,
  author  = {Joelle Pineau and Philippe Vincent-Lamarre and Koustuv Sinha and Vincent Larivi{\`e}re and Alina Beygelzimer and Florence d'Alch{\'e}-Buc and Emily Fox and Hugo Larochelle},
  title   = {Improving Reproducibility in Machine Learning Research (a Report from the {NeurIPS} 2019 Reproducibility Program)},
  journal = {Journal of Machine Learning Research},
  volume  = {22},
  number  = {164},
  pages   = {1--20},
  year    = {2021},
  url     = {https://www.jmlr.org/papers/v22/20-303.html}
}

@article{LopezPerez2023SmartSeville,
  author  = {Mar{\'i}a Eugenia L{\'o}pez-P{\'e}rez and Mar{\'i}a Eugenia Reyes-Garc{\'i}a and Mar{\'i}a Eugenia L{\'o}pez-Sanz},
  title   = {Smart Mobility and Smart Climate: An Illustrative Case in {Seville}, {Spain}},
  journal = {International Journal of Environmental Research and Public Health},
  volume  = {20},
  number  = {2},
  pages   = {1404},
  year    = {2023},
  doi     = {10.3390/ijerph20021404}
}

@article{Laso2022ElasticAnalytics,
  author  = {Sergio Laso and Javier Berrocal and Pablo Fern{\'a}ndez and Jos{\'e} Mar{\'i}a Garc{\'i}a and Jos{\'e} Garc{\'i}a-Alonso and Juan M. Murillo and Antonio Ruiz-Cort{\'e}s and Schahram Dustdar},
  title   = {Elastic Data Analytics for the Cloud-to-Things Continuum},
  journal = {IEEE Internet Computing},
  volume  = {26},
  number  = {6},
  pages   = {42--49},
  year    = {2022},
  doi     = {10.1109/MIC.2021.3138153}
}

@inproceedings{Keranen2009ONE,
  author    = {Ari Ker{\"a}nen and J{\"o}rg Ott and Teemu K{\"a}rkk{\"a}inen},
  title     = {The {ONE} Simulator for {DTN} Protocol Evaluation},
  booktitle = {Proceedings of the 2nd International Conference on Simulation Tools and Techniques (SIMUTools)},
  pages     = {1--10},
  year      = {2009},
  doi       = {10.4108/ICST.SIMUTOOLS2009.5674}
}

@article{laso2026evaluating,
  title={Evaluating data analytics architectures performance: a comparative study on the continuum},
  author={Laso, Sergio and Berrocal, Javier and Fern{\'a}ndez, Pablo and Ruiz-Cort{\'e}s, Antonio and Murillo, Juan Manuel and Dustdar, Schahram},
  journal={Journal of Cloud Computing},
  year={2026},
  publisher={Springer},
  doi     = {10.1186/s13677-026-00960-z}
}

\end{document}